\documentclass[10pt,longbibliography,prx,aps,twocolumn]{revtex4-2}
\usepackage{graphicx}
\usepackage{xcolor}
\usepackage{braket}
\usepackage{amsmath,bm}
\usepackage{enumitem}
\usepackage{mathrsfs} 
\usepackage{multirow}
\usepackage{booktabs}
\usepackage[normalem]{ulem}
\usepackage[colorlinks]{hyperref}
\usepackage[dvipsnames]{xcolor}

\DeclareMathOperator{\rnk}{rnk}

\begin{document}
\title{Non-Hermitian multimode interferometry}
\author{Subhajyoti Bid}
\affiliation{School of Physics and Astronomy, Lancaster University, Lancaster, LA1 4YB, United Kingdom}
\author{Henning Schomerus}
\affiliation{School of Physics and Astronomy, Lancaster University, Lancaster, LA1 4YB, United Kingdom}
\date{\today}
\begin{abstract}
Interferometers form a cornerstone of precision measurement, leveraging wave superposition and phase coherence to convert minute physical perturbations into
measurable intensity variations. 
Here, we establish how paradigmatic interferometric architectures, the canonical Michelson and Mach-Zehnder interferometers, can be endowed with entirely new operating principles by replacing conventional beam splitters with judiciously designed non-Hermitian resonators. 
To uncover how non-Hermitian physics can assist in interferometry, we subject these canonical interferometers to a modern non-Hermitian symmetry analysis. 
A fundamentally transformed interference landscape then arises when the resonators and interference pathways collectively induce partially deficient spectral degeneracies, mathematically characterized as multimode versions of exceptional points. 
Key characteristics of these \emph{exceptional interferometers} are nonanalytic destructive interference conditions, resulting in distinct bright-fringe and dark-fringe operating regimes organized by non-Hermitian  winding and braiding topology.
By revealing how exceptional-point-assisted interference reshapes 
the canonical geometries, our findings transfer 
central paradigms of non-Hermitian spectroscopy into the interferometric setting,
and establish exceptional-point interferometry as a new platform-independent paradigm for general-purpose precision sensing, applicable from on-chip photonics to macroscopic observatories.\end{abstract}

\maketitle

\section{Introduction}

Interference has long provided one of the most sensitive ways to 
reveal the properties of 
physical systems. 
Early demonstrations by Young, Fresnel, and Arago \cite{young,fresnel1816memoire,arago1819action} showed that coherent wave superposition encodes subtle phase information, an insight that transformed light into a precision probe of nature. Foundational experiments, from Young’s double slit fringes to Arago’s purpose built instruments, established interference as a tool for detecting perturbations far below direct measurement thresholds. These ideas culminated in the seminal Michelson and Mach-Zehnder geometries \cite{Michelson1887On,nolte2023}, which elevated interferometry into a cornerstone of precision science \cite{hecht2017optics}. Further architectures such as Sagnac, Fabry-Perot, Ramsey, and Hanbury Brown-Twiss broaden this scope to high-resolution spectroscopy up to the quantum limit \cite{scully2014quantum,meystre2014elements}. Today, interferometers underpin applications from metrology and spectroscopy to optical coherence tomography, atomic clocks, integrated photonic sensing, accelerometry, and gravitational wave detection, all exploiting the extraordinary sensitivity that arises from the controlled interference between multiple propagation pathways.

Over the recent decade, non-Hermitian physics has matured into a rigorous framework for analyzing open photonic systems where gain, loss, and radiative coupling play dedicated roles \cite{microjan2, ElGanainy2018}. A hallmark of this regime is the emergence of exceptional points (EPs), defective spectral degeneracies where both resonance frequencies and the corresponding spatial mode profiles coalesce \cite{Kato66,Heiss2012,Miri2019}.
Associated with a nontrivial parameter-space topology \cite{Berry2004,Bergholtz2021}, 
which is directly probed in
chiral mode conversion \cite{Doppler2016,Ghosh2016}, 
these non-Hermitian degeneracies induce a host of unconventional phenomena, 
among which an anomalously enhanced, nonanalytic sensitivity to external perturbations 
\cite{microjan1,Chen2017, Hodaei2017} is of particular metrological relevance \cite{Wiersig:20}. This mechanism promises dramatic sensitivity gains, driving EP realizations in diverse architectures and platforms \cite{Peng2014,Xu2016,Zhang2017,Zhong2019,Li2023}. 
Selectively addressing specific input-output channels further enables the tuning of the effective sensitivity exponent \cite{Zhong2020,Kullig:23}, balancing enhancement against robustness to disorder and paving
the way for next-generation non-Hermitian sensor technologies.
Viewed through the lens of interferometry, these defective spectral degeneracies act as singular points where multiple scattering pathways collapse onto a fully degenerate eigenmode manifold, thereby amplifying the response to symmetry-breaking perturbations. Granting full interferometric access to multiple scattering
pathways should provide an additional layer of control over this observable response, and enable genuinely interferometric operating principles that utilize a richer, partially degenerate eigenmode manifold. Consequently, non-Hermitian multipath architectures stand out as an untapped resource for precision interferometry.

\begin{figure*}[t]
    \centering
        \includegraphics[width=0.9\linewidth]{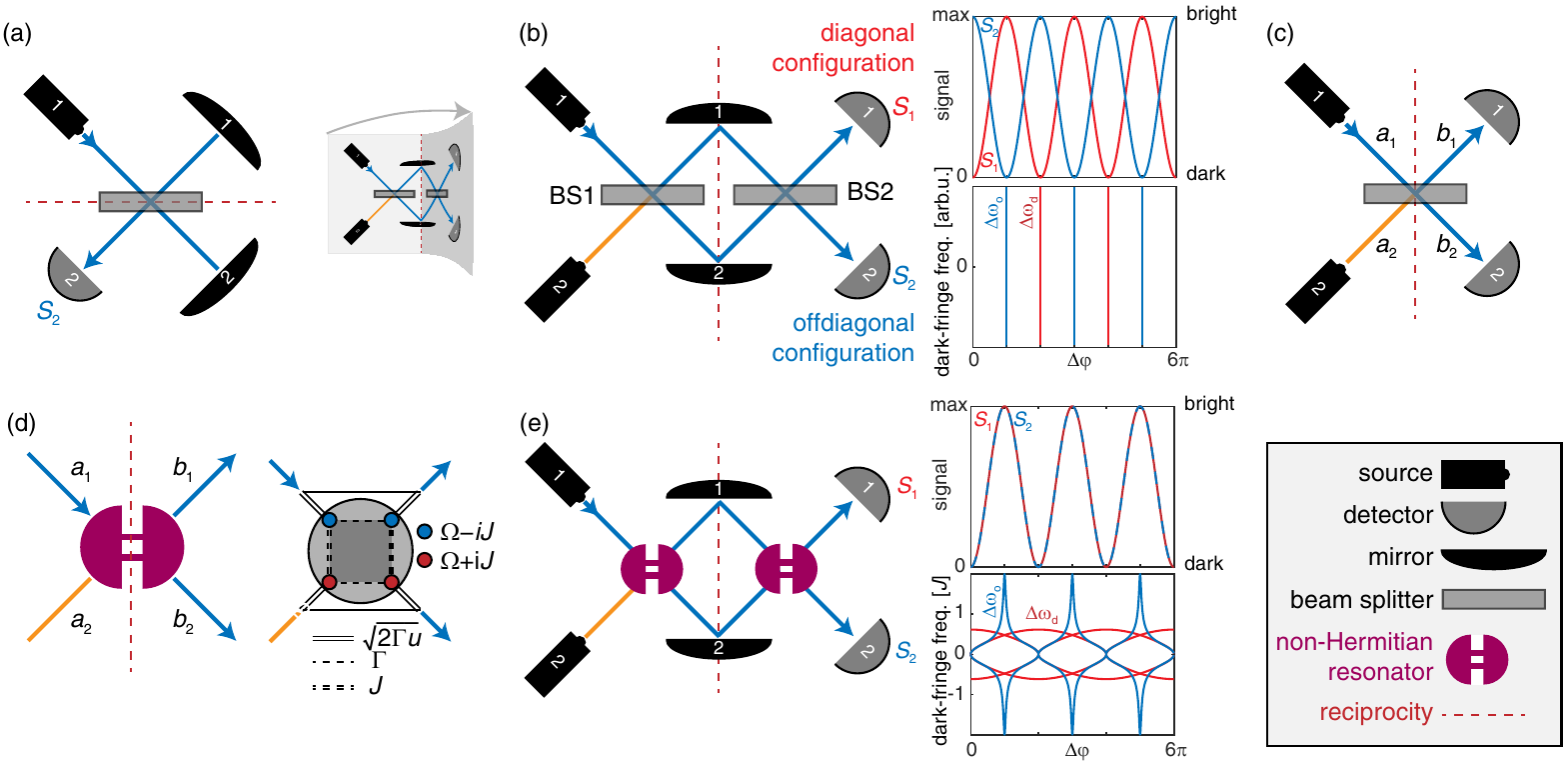}
    \caption{Exceptional interferometer principle. 
    (a) Conventional Michelson interferometer geometry, in which a beam is split and recombined by a 50:50 beam splitter. 
     The red dashed line indicates  Lorentz reciprocity,
   which interchanges source and detector. 
    (b) Mach-Zehnder interferometer geometry, interpreted in our symmetry analysis as an unfolded Michelson interferometer. 
   Insets: The two detector configurations display complementary interference signals $S_1$, $S_2$ that depend on the phase difference $\Delta\phi$ accumulated in the arms. At any frequency, 
   dark fringes appear at a fixed phase difference $\Delta \phi=0\pmod{2\pi}$ or $\pi \pmod{2\pi}$, respectively. 
    (c) Input and output channels of a conventional 50:50 beam splitter. (d) Non-Hermitian resonant beam splitter (left) and its minimal realization based on four internally coupled components with gain and loss (right, see Appendix~\ref{app:nhresonator} for further details), operating at a defective spectral degeneracy (a single-mode exceptional point, EP).  
   (e) Exceptional interferometer, in which
 the multipath interference induces a collective multimode EP 
    whose internal structure is highly sensitive to the interference conditions. Insets: The output signals are synchronized (upper panel, MEP critical working point), and display a nonanalytic phase dependence of the dark-fringe frequency at the diagonal and off-diagonal critical working point (lower panel), shown here for $\Gamma/J=0.1$ (see text for definitions of these parameters and working points).   
}
    \label{fig:main_model}
\end{figure*}

Here, we present a non-Hermitian interferometry paradigm where multimode EPs arise as a collective phenomenon from the intrinsic multipath geometry, thereby establishing a fundamentally new sensing principle. We develop this paradigm concretely through non-Hermitian extensions of Michelson and Mach-Zehnder interferometers
in which the central optical elements, conventionally provided by 50:50 beam splitters, are replaced by non-Hermitian resonators (see Fig.~\ref{fig:main_model}), and show that these familiar geometries acquire dramatically modified characteristics when operated near-resonantly at their collective multimode EP.
The detailed quantitative analysis of the ensuing
interferometer signals uncovers distinct bright- and dark-fringe operating regimes, identifies
critical working points, and reveals their connection to non-Hermitian winding and braiding topology. As the required components are standard in existing non-Hermitian technologies 
\cite{Ruter2010,microjan2,Van2016Microring,QuirozJuarez:19,Zhong2019,
Zhong2020,Soleymani2022,Kullig:23,Li2023,On2024}, our findings provide a complete design pathway
to exceptional interferometry on a wide range
of platforms.

The paper is organized as follows: In the background Sec.~\ref{sec:background}, we establish the symmetries of conventional Michelson and Mach-Zehnder interferometers.  The design of the exceptional interferometer utilizing the non-Hermitian generalization of these symmetries is presented 
in Sec.~\ref{sec:eidesign}, while the detailed analysis of the operating regimes and working points is presented in Sec.~\ref{sec:eioperation}. Our conclusions and outlook are given in Sec.~\ref{sec:outlook}. Some technical details of the derivations and additional numerical results are collected in the Appendices.

\section{Background: Symmetry analysis of conventional interferometers\label{sec:background}}
To place exceptional interferometers into a wider context, we first subject classical architectures to a simple symmetry analysis and identify the main constraints imposed by the conventional Hermitian optical elements. 
We start with the Michelson interferometer in its canonical form, which splits an incident field into two arms using a 50:50 beam splitter, reflects each component from two mirrors, and recombines them by the same beam splitter to generate an interference signal at the detector
[see Fig.~\ref{fig:main_model}(a)]. Comprising linear, passive, and time-invariant elements, this geometry obeys Lorentz reciprocity: exchanging the input and output ports leaves the interferometric response unchanged. 
This reciprocal symmetry ensures that the phase accumulation is independent of the propagation direction, 
so that the interference signal depends exclusively on the 
relative optical phase difference $\Delta \phi$ accumulated in the two arms.
The resulting interference fringes are highly sensitive to changes in $\Delta \phi$,
enabling the Michelson interferometer to measure length variations with precision of the order of the optical wavelength $\lambda$. The equal intensity splitting between the two arms maximizes the fringe contrast, establishing complete destructive interference ($\Delta \phi \equiv \pi \mod 2\pi$) as a robust operating regime driven by the precise cancellation of two amplitudes that are exactly out of phase.

Unfolding the Michelson geometry yields the canonical Mach-Zehnder architecture shown in Fig.~\ref{fig:main_model}(b), in which light propagates in one direction. Eliminating the mirror-based retro-reflection necessitates the use of two beam splitters (BS1 and BS2), which now divide and recombine the fields in a fully guided geometry.
Identical beam splitters restore an effective Lorentz reciprocity that preserves the formal equivalence to the Michelson setup.  
However, as indicated in the figure, this unfolded geometry allows for two nonequivalent 
\emph{diagonal} and \emph{off-diagonal} source-detector configurations, whose 
respective interference signals are strictly complementary due to intensity conservation in these nondissipative systems (see insets). We refer to these two configurations as the
\emph{diagonal configuration}
and the \emph{off-diagonal configuration}.
Interpreted through this unfolded geometry, the Michelson interferometer realizes the off-diagonal configuration.

Mathematically, we can express these symmetries and interference conditions in terms of the $2\times2$ transmission matrices $t_1$ and $t_2$ for the beam splitters, which connect the incoming amplitudes $a_{1,2}$ on one side to the outgoing amplitudes $b_{1,2}$ on the other side [see Fig.~\ref{fig:main_model}(c)], 
as well as a diagonal transmission matrix $t_a$ for the propagation through the arms. The diagonal and off-diagonal elements $t_{ij}$ of the total transmission matrix $t=t_2t_at_1$ then dictate the signal propagation for all source-detector combinations in the Mach-Zehnder setup, with the canonical Michelson setup described by the lower left element $t_{21}$. Lorentz reciprocity enforces $t_1=t_2^T$ and $t_a=t_a^T$. For identical beam splitters ($t_1=t_2$), $t=t^T$, and in the lossless case $t$ is unitary, which mathematically defines the Hermitian setting. The bright-fringe and dark-fringe operating conditions are then fundamentally governed by symmetry: equivalent arms impose $t=\sigma_x t \sigma_x$ where $\sigma_x$ is the Pauli $x$ matrix, which is compatible with constructive and destructive interference in the off-diagonal and diagonal configurations, respectively.

These constraints are verified using standard 50:50 beam splitters, defined in a symmetric basis by  \cite{Loudon} 
\begin{equation}
    t_1=t_2=\frac{1}{\sqrt{2}}\begin{pmatrix} 1 & i \\ i & 1 \end{pmatrix}.
\end{equation} 
These beam splitters do not discriminate between the pairs of incoming and outgoing modes, whereby $t_1=\sigma_x t_1 \sigma_x$, $t_2=\sigma_x t_2 \sigma_x$. 
The propagation of the fields through the arms is described by a diagonal matrix
$t_a=\mathrm{diag}(\kappa_1, \kappa_2)$, where the complex coefficients $\kappa_i$ capture both the cumulative effective optical path length (phase) and field attenuation (amplitude), including the reflectivity of the mirrors and gain and loss in the propagation through the arms. The total transmission matrix then takes the simple standard form
\begin{equation}
t=\frac{1}{2}\begin{pmatrix} \kappa_1-\kappa_2&i(\kappa_1+\kappa_2)\\i(\kappa_1+\kappa_2)&\kappa_2-\kappa_1 \end{pmatrix}.
\end{equation}
Parameterizing the path-length difference via 
\begin{equation}
\label{eq:signalparametrisation}
    \kappa_2/\kappa_1=c e^{i\Delta\phi}\qquad (c \geq 0)
\end{equation}
 yields the characteristic interference signals 
\begin{equation}
\begin{aligned}
S_1\equiv\left|\frac{t_{11}}{\kappa_1}\right|^2=\frac{1+c^2-2c\cos\Delta\phi}{4}&
&&\mbox{(diagonal)},
\\
S_2\equiv\left|\frac{t_{21}}{\kappa_1}\right|^2
=\frac{1+c^2+2c\cos\Delta\phi}{4}&  &&\mbox{(off-diagonal)},
\end{aligned}
\label{eq:conv_signals}    
\end{equation}
which display complementary sinusoidal patterns [see upper inset of Fig.~\ref{fig:main_model}(b)]. 
As illustrated in the lower inset, complete destructive interference occurs at $c=1$ and under the standard
frequency-independent phase conditions
\begin{equation}
\begin{aligned}
\Delta\phi=0 \pmod{2\pi}& && \mbox{(diagonal)},
\\
\Delta\phi=\pi \pmod{2\pi}& && \mbox{(off-diagonal)},
\end{aligned}
\label{eq:phaseconditions}    
\end{equation}
These conventional conditions coincide with the symmetry constraints $\sigma_x t\sigma_x = \pm t$, which are realized for equivalent arms with $\kappa_1=\kappa_2$, and $\pi$-shifted signals with $\kappa_1=-\kappa_2$,
respectively.

\begin{figure*}
    \centering
    \includegraphics[width=\linewidth]{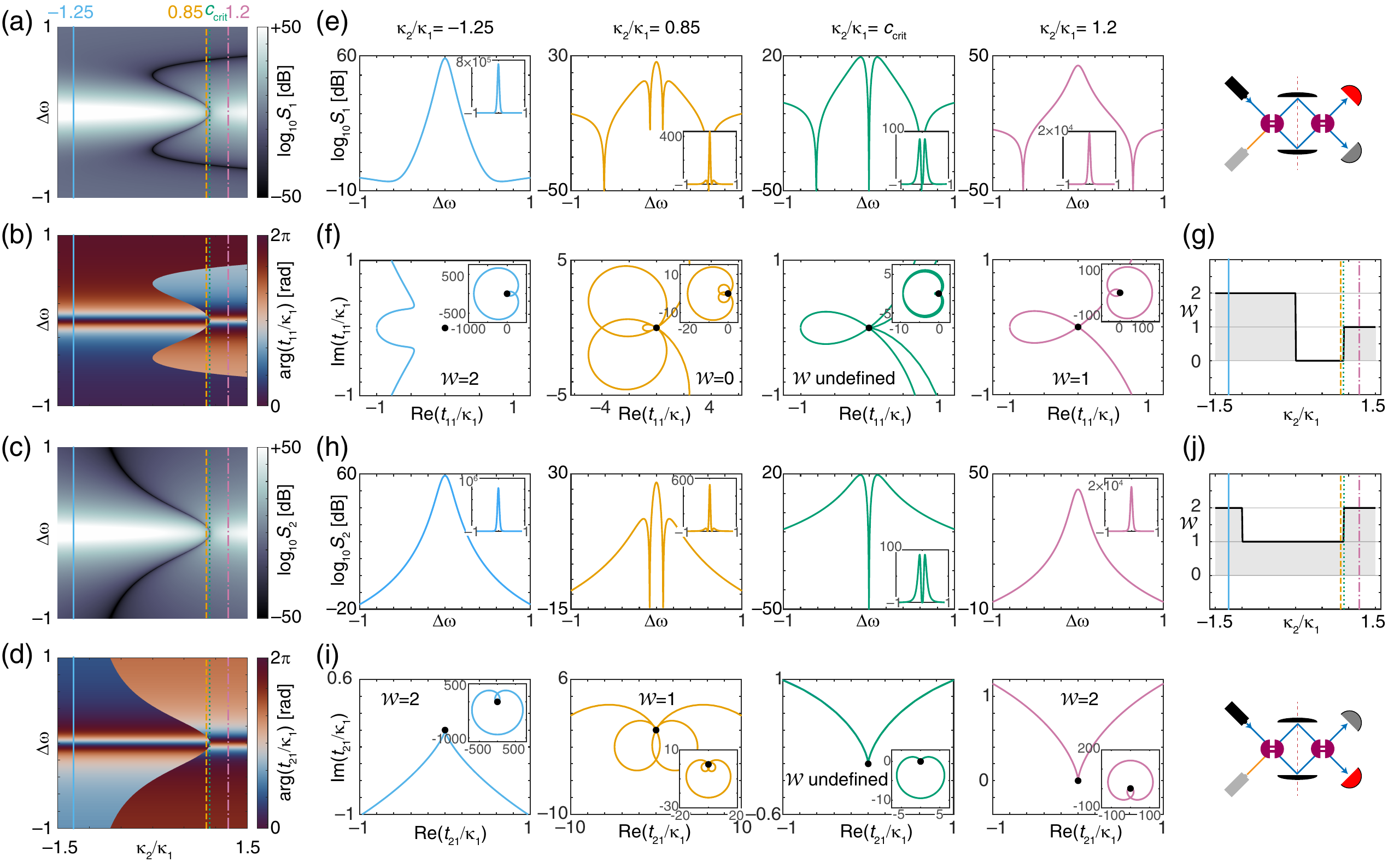}
    \caption{Interference landscape and topologically distinct bright-fringe operating regimes. All panels are evaluated for a non-Hermitian resonant beam splitter with $J=1$, $\Gamma=0.1$, $\gamma=0$. Panels (a,b,e,f,g) are for the diagonal source-detector configuration, the remaining panels are for the off-diagonal configuration.
    (a,c) Frequency-resolved signal intensities $|t_{11}/\kappa_1|^2$ and $|t_{21}/\kappa_1|^2$ on a logarithmic dB scale versus effective arm-propagation coefficient ratio $\kappa_2/\kappa_1$ and detuning frequency $\Delta\omega$. The bright resonance near $\Delta\omega=0$ is disrupted by a sharp dip induced by the multimode EP at $\kappa_2/\kappa_1=1$, resulting in a crossing with the dark-fringe curve where the signal vanishes. (b,d) Corresponding signal phases $\mathrm{arg}\,t_{11}/\kappa_1$ and $\mathrm{arg}\,t_{21}/\kappa_1$, in which the resonance shows up via rapid winding and the dark-fringe curve via a $\pi$ jump. (e,h) Frequency-resolved signal intensities at fixed values of $\kappa_2/\kappa_1$, indicated in panels (a-d) as vertical lines.
    (f,i) Frequency-dependent signal phase in an Argand diagram, providing a classification of different bright-fringe regimes in terms of a quantized topological  winding number, shown in  (g,j). 
    } 
    \label{fig:main_results}
\end{figure*}

This setting establishes the familiar Hermitian interference landscape, in which unitarity of the transmission matrices constrains the accessible interference signals.
The symmetry analysis emphasizes two hallmarks of these constraints, the complementary of the interference signals and the strict separation of bright-fringe and dark-fringe conditions.
This  provides us with guidance
for designing controlled non-Hermitian elements that transcend these limitations, paving the way for the exceptional interferometer principle described next.

\section{Exceptional interferometer design\label{sec:eidesign}}
To overcome the conventional interference constraints emphasized by the preceding symmetry analysis, we now turn to the design of an interferometer that utilizes the non-Hermitian sensing characteristics of exceptional points. Concretely, we devise these exceptional interferometers by replacing the conventional 50:50 beam splitters with non-Hermitian optical elements designed
to induce a collective generalized multimode EP. The goal is to equip the interference signal with a resonant response that drastically depends on the arm propagation coefficients $\kappa_1$ and $\kappa_2$, resulting in highly characteristic noncomplementary signals and 
bright and dark fringe interference conditions that are constrained by non-Hermitian symmetry and topology. 
We first focus on designing the system and explaining the underlying operating principle, a symmetry-induced transition of the EP from a single-mode to a multimode character that is intimately tied to the collective multi-path geometry of the interferometer. 

\subsection{Beam-splitter design}
To achieve the desired functionality, 
we utilize non-Hermitian two-mode resonators described by an effective non-Hermitian Hamiltonian 
\begin{equation}
H_\mathrm{eff}= \begin{pmatrix} \Omega -iJ&J\\J&\Omega+iJ \end{pmatrix},    
\end{equation}
where $\Omega\equiv \Omega_0-i\gamma$ are bare resonator frequencies that can be complex, while $J$ represents the mutual coupling of the corresponding internal modes.
This minimal model Hamiltonian preserves reciprocity, $H_\mathrm{eff}=H_\mathrm{eff}^T$, guaranteeing that the system can be realized with standard optical components (see Fig.~\ref{fig:main_model}(d) and Appendix \ref{app:nhresonator}, as well as the overview of existing technologies in the conclusions). 
The Hamiltonian possesses a two-fold degenerate eigenvalue $\Omega$ but only a single eigenvector $(1,i)^T$, hence at this stage still describing a conventional single-mode EP.

Using standard input-output theory (see Appendix~\ref{app:nhresonator}),
the transmission matrix for each passage through the resulting non-Hermitian beam splitter is given by 
\begin{equation}
t_1=t_2=i\openone+2\Gamma[(\omega+i\Gamma)\openone -H_\mathrm{eff}]^{-1},    
\end{equation}
where $\Gamma$ is the coupling strength of the resonator to the interferometer arms. Introducing the frequency detuning $\Delta \omega=\omega-\Omega_0+i\gamma$, we can write these transmission matrices explicitly as 
\begin{equation}
t_1=t_2=i\frac{\Delta\omega -i\Gamma}{\Delta\omega +i\Gamma}\openone + \frac{2\Gamma J}{(\Delta\omega +i\Gamma)^2}\begin{pmatrix} -i&1\\1&i \end{pmatrix},    
\end{equation}
verifying that the single-mode EP induces a second-order super-Lorentzian transmission resonance at $\omega=\Omega$ \cite{Yoo2011,Hashemi2022}.

\subsection{Collective multimode exceptional point}
The key idea of the exceptional interferometer, shown in Fig.~\ref{fig:main_model}(e), is to 
combine the double-passage through the resonant beam splitters with the propagation through the interferometer arms into an emergent collective multimode EP, and exploit the resulting nonanalytic bright-fringe and dark-fringe interference conditions. 
This is borne out through the mathematical structure of the total transmission matrix $t=t_2 t_a t_1$ of the complete interferometer, 
\begin{widetext}
\begin{align}
    t =   
    -\left( \frac{\Delta\omega - i\Gamma}{\Delta\omega + i\Gamma} \right)^2 \begin{pmatrix} \kappa_1 & 0 \\ 0 & \kappa_2 \end{pmatrix} 
    + \frac{2i\Gamma J(\Delta\omega - i\Gamma)}{(\Delta\omega + i\Gamma)^3} \begin{pmatrix} -2i\kappa_1 & \kappa_1+\kappa_2 \\ \kappa_1+\kappa_2 & 2i\kappa_2 \end{pmatrix} 
    + \frac{4\Gamma^2 J^2(\kappa_1-\kappa_2)}{(\Delta\omega + i\Gamma)^4} \begin{pmatrix} -1 & -i \\ -i & 1 \end{pmatrix},
\label{eq:main_result}
\end{align}
\end{widetext}
which combines the resonant response of the non-Hermitian beam splitters with the interference of the signal propagation through the interferometer arms. 
The terms in Eq.~\eqref{eq:main_result} describe resonant behavior of different orders, leading to a strongly enhanced signal near resonance, $\Delta \omega=0$. 
As its key property, the contribution associated with a fourth-order super-Lorentzian peak  generally dominates, but vanishes identically for $\kappa_1=\kappa_2$.   
This suppression of the quartic term leaves the next-leading contribution as the dominant feature near resonance, giving rise to a third-order super-Lorentzian response,
while the resonance condition remains unchanged.

\begin{figure*}[t]
\centering
\includegraphics[width=0.95\linewidth]{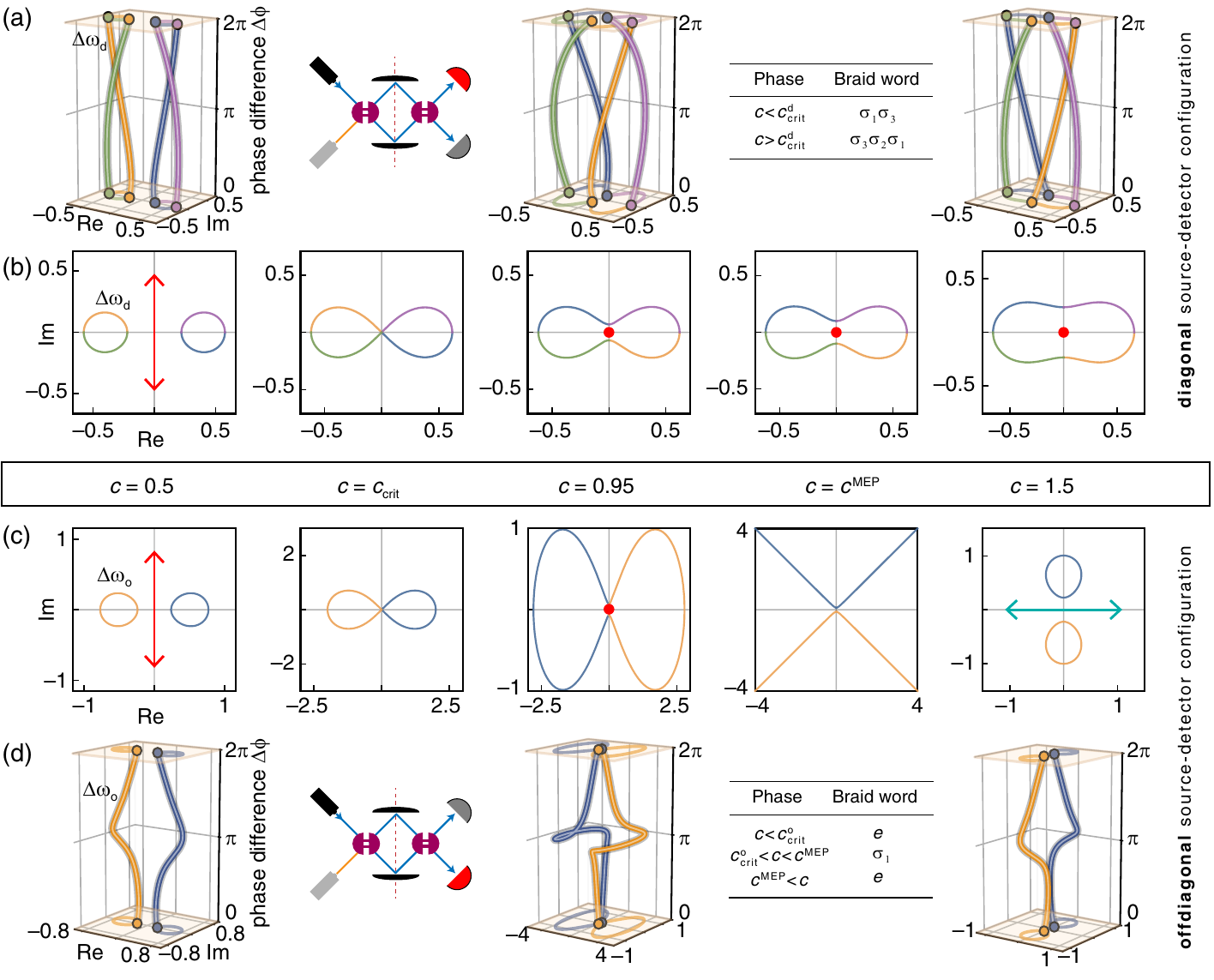}
\caption{Topological dark-fringe operating regimes.
(a,b) Trajectories of the dark-fringe detuning frequencies \eqref{eq:wp_modes} in the complex plane and corresponding braiding diagrams in the diagonal source-detector configuration,
for variable $\Delta\phi \in [0, 2\pi]$ and fixed value of the arm asymmetry parameter $c=|\kappa_2/\kappa_1|$.
(c,d) Corresponding results in the off-diagonal source-detector configuration.
The sequences capture transitions from an imaginary line gap via a gap-closing singularity ($c = c^{o/d}_\mathrm{crit}$) into a point-gap topology. At $c=1$, the off-diagonal dark-fringe detuning frequencies run off to infinity, resulting in an additional transition into a real line-gap phase. Insets list the braid words for the different phases (see Appendix~\ref{app:braiding} for the detailed derivation).
In all panels, $J=1$ and $\Gamma=0.1$.}
\label{fig:topological_evolution}
\end{figure*}

\subsection{Multimode exceptionality}
The drastically restructured resonant response at $\kappa_1 = \kappa_2$ is a collective property of all components in the exceptional interferometer. To determine its mathematical origin, we construct an effective collective Hamiltonian that describes the complete system, which will reveal an abrupt transition from a single-mode EP at $\kappa_1 \neq \kappa_2$ to a multimode EP at $\kappa_1 = \kappa_2$. This is achieved by bringing the total transmission matrix  \eqref{eq:main_result}
into the mathematically identical form
\begin{align}
t=-KK^T+4i\Gamma K[(\Delta \omega+i\Gamma)\openone-N]^{-1}K^T,
\label{eq:teff}
\end{align}
involving an effective external coupling matrix
\begin{align}
K=\begin{pmatrix}
    \sqrt{\kappa_1} & 0 & 0 & 0
    \\
    0 & \sqrt{\kappa_2} & 0 & 0 
\end{pmatrix},
\label{eq:keff}
\end{align}
and the effective collective Hamiltonian
\begin{align}
N=\begin{pmatrix}
    -i(\Gamma+J) & \kappa_{+} J & \Gamma & -iJ \kappa_{-}
    \\
    \kappa_{+} J & -i(\Gamma-J) & -iJ \kappa_{-} & -\Gamma \\
    \Gamma & -iJ \kappa_{-} & i(\Gamma-J) & -\kappa_{+} J \\
    -iJ \kappa_{-} & -\Gamma & -\kappa_{+} J & i(\Gamma + J)
\end{pmatrix},
\label{eq:neff}
\end{align}
with $\kappa_{\pm}=(\kappa_1\pm \kappa_2)/(2\sqrt{\kappa_1\kappa_2})$.

The effective collective Hamiltonian $N$ subsumes the spectral characteristics of the complete interferometer, comprising the two resonant beam splitters as well as the two arms. This Hamiltonian obeys reciprocity in the form $N=N^T$, 
complying with the general symmetry requirements and realizability by standard optical components.
The property that establishes the exceptional multimode functionality is the nilpotency of the effective Hamiltonian, which obeys $N^4=0$ in general, while in addition $N^3=0$ for equivalent arms, $\kappa_1=\kappa_2$. These features dictate a change of the nature of the exceptional point: 
While the resonance remains four-fold degenerate for all values of $\kappa_1$ and $\kappa_2$,  a second eigenmode appears for $\kappa_1=\kappa_2$. Therefore, the spectral degeneracy changes from a single-mode exceptional point to a multimode exceptional point (technically, a 3+1 two-mode exceptional point, see Appendix~\ref{app:mmep} for a detailed exposition of the stated features of $N$).

The resulting restructured resonant response of the exceptional interferometer is recovered from the exact geometric-series expansion \cite{Heiss2015,Hashemi2022,Wiersig2023,bid2024uniform}
\begin{align}
t=-KK^T+4i\Gamma \sum_{l=0}^3\frac{KN^{l}K^T}{(\Delta \omega+i\Gamma)^{l+1}}
\label{eq:teff2}
\end{align}
of 
Eq.~\eqref{eq:teff}. As indicated, this exact expansion generally terminates at $l=3$, but
terminates at $l=2$ when $N^3=0$, hence, precisely when the multimode EP is realized.

This resonance mechanism constitutes a unique spectral interference phenomenon, distinct, e.g., from Fano resonances, Dicke resonances, and bound states in the continuum.
As we establish in the following, this mechanism provides versatile pathways for reconfigurable non-Hermitian interferometry in a variety of operating regimes, distinguished by non-Hermitian winding and braiding topology.

\section{Operating regimes and working points
\label{sec:eioperation}}

Choosing to operate any interferometer near constructive or destructive interference (bright or dark fringes) dictates its sensitivity
and tolerance to noise, and results in fundamental
differences in the measured signals.
In particular, dark-fringe operation benefits from an objectively well-defined working point independent of input intensity and detector efficiency, but often requires active stabilization and compensation for dispersion.
While such tradeoffs also apply to the exceptional interferometer, its operation is enriched by the multimode interference and 
noncomplementarity 
of the diagonal and off-diagonal
signals induced by the non-Hermitian resonators.
We therefore establish a variety of distinct
bright-fringe and dark-fringe operating regimes, organized by non-Hermitian winding and braiding topology of the working points. 

Practically, the different regimes  are attained by the choice of interferometer parameters and working points. 
To establish these regimes, we first give an overview of the interference landscape in the principal parameter space, spanned by the complex detuning frequency $\Delta\omega$ and the arm-propagation coefficient ratio $\kappa_2/\kappa_1$. We then discuss resonant bright-fringe operation, showing that different regimes are distinguished by the winding properties of the interference signal. Next, we determine the dark-fringe conditions, leading to operating regimes that are distinguished by braiding of the dark-fringe detuning frequency $\Delta \omega_d$. Finally, the intersection of both conditions results in mixed operating modes,
coinciding with the critical values of the topological phase transitions, where the multimode nature of the EP induces non-analytical frequency shifts.

\subsection{Overview of the interference landscape}

All operating regimes are based on the spectral richness that the multimode EP imprints onto the interference signals in the different source-detector configurations, as determined by the total transmission matrix \eqref{eq:main_result}.
In conventional interferometers, the unattenuated diagonal and off-diagonal signals are complementary to each other, i.e., add up to a constant total intensity that equals the intensity of the source.
The exceptional interferometer departs from this behavior because of the dissipative nature of  non-Hermitian resonators.
Beyond this, the signals exhibit a systematic resonance structure whose order depends on the conditions of constructive and destructive interference, reflecting the multimode reconfiguration of the EP as established in the design section. 

For guidance through the resulting interference landscape,
Fig.~\ref{fig:main_results} provides an overview of these signals as a function of the detuning frequency $\Delta\omega$ and the arm-propagation coefficient ratio $\kappa_2/\kappa_1$, both taken real, and the coupling ratio $\Gamma/J = 0.1$ of the components in the non-Hermitian resonant beam splitter fixed to a representative value in which all qualitative features are clearly visible.
The scaled diagonal signal intensity $S_1=\vert{}t_{11}/\kappa_1\vert{}^2$, shown on a logarithmic dB scale in Fig.~\ref{fig:main_results}(a), peaks near the resonance line $\Delta\omega=0$,
but dips
sharply around the multimode EP at $\kappa_2/\kappa_1=1$. This dip extends to finite detuning as a continuous dark-fringe curve $S_1=0$ where the signal exactly vanishes. Correspondingly, the phase $\mathrm{arg}(t_{11}/\kappa_1)$ [Fig.~\ref{fig:main_results}(b)] winds rapidly near resonance and undergoes a discrete $\pi$ jump upon crossing the dark-fringe curve.
Analogous features show up for the off-diagonal configuration, displayed in  Fig.~\ref{fig:main_results}(c,d) for the corresponding scaled signal $S_2=|t_{21}/\kappa_1|^2$. 
These bright and dark features, governed by the resonance condition $\Delta\omega=0$ and the dark-fringe curve $S_i=0$, form the basis for the specific operating regimes that we describe next.

\subsection{Bright-fringe operation and winding topology}

We now describe the interferometer operation around resonance, $\Delta\omega=0$, where the signal is generally bright. Using the arm-propagation parameterization \eqref{eq:signalparametrisation}
and expanding the signals
for small $\Gamma/J$, the leading-order signals 
\begin{equation}
|t_{11}/\kappa_1|^2\approx |t_{21}/\kappa_1|^2\approx\frac{16J^4}{\Gamma^4}(1+c^2-2c\cos\Delta\phi)    
\end{equation}
in both source-detector configurations share the phase dependence of a conventional interferometer in the diagonal configuration, see Eq.~\eqref{eq:conv_signals}. However, because both configurations exhibit identical phase dependence, they are synchronized rather than complementary, and resonantly enhanced. 

A further distinction arises when we account for the lineshape of the frequency-resolved intensity around the resonance. 
As shown in Fig.~\ref{fig:main_results}(e,h), detailed resonance features arise from the interplay of the resonance condition with the dark-fringe curve,
which imprints a distinct side-band structure. The features are further elucidated by the frequency-dependence of the complex signal amplitude, visualized as trajectories in Argand diagrams in Fig.~\ref{fig:main_results}(f,i). 
Topological transitions occur when the trajectory passes through the origin, representing 
the formation or destruction of two dark-fringe working points.

We classify the global signal topology using a quantized winding number $\mathcal{W}$ [Fig.~\ref{fig:main_results}(g,j)], determining $\mathcal{Z}=4-2\mathcal{W}$ dark-fringe working points embedded in the frequency-resolved signal (Appendix~\ref{app:polecomp}).
The winding number undergoes abrupt steps at critical thresholds
\begin{equation}
\begin{aligned}
\frac{\kappa_2}{\kappa_1}&=\left(1-\frac{\Gamma}{2J}\right)^2\equiv c^d_\mathrm{crit}    &&\mbox{(diagonal)},\quad
\\
\frac{\kappa_2}{\kappa_1}&=\frac{1-\frac{\Gamma}{2J}}{1+\frac{\Gamma}{2J}}\equiv c^o_\mathrm{crit}     &&\mbox{(off-diagonal)}
,
\end{aligned}
\label{eq:crit}
\end{equation}
which approach the multimode EP point 
\begin{equation}
\kappa_2/\kappa_1 = 1 \equiv c^\mathrm{MEP} \end{equation}
in the weak-coupling limit $\Gamma\ll J$.
These transitions directly dictate the observable resonant lineshape:
a unit step $|\Delta \mathcal{W}|=1$  marks the emergence or annihilation of a pair of dark-fringe working points embedded in the frequency-resolved signal.

We will next describe operation regimes that are directly based on these dark-fringe working points, and then return to establish mixed bright-dark mode interferometer principles that emerge at the critical points themselves.

\subsection{Dark-fringe operation and braiding topology}

We now specify operation regimes of the exceptional interferometer
that are based on the dark-fringe condition $S_i=0$, corresponding to dark-fringe operation under conditions of complete destructive interference. The conditions depend on the source-detector configuration, and are obtained by equating the corresponding total transmission coefficient from Eq.~\eqref{eq:main_result} to zero. This yields complete destructive interference at complex frequency detuning 
\begin{equation}
\begin{aligned}
(\Delta\omega_d^\pm)^2 &\equiv 2\Gamma J\left(1 \pm \sqrt{\frac{\kappa_2}{\kappa_1}}\right) -\Gamma^2&&\mbox{(diagonal)}, \\
(\Delta\omega_o)^2 &\equiv 2\Gamma J\frac{\kappa_1-\kappa_2}{\kappa_1+\kappa_2} - \Gamma^2 &&\mbox{(off-diagonal)}.
\end{aligned}
\label{eq:wp_modes}
\end{equation}
These working points can be attained by adjusting any two independent parameters, including the detuning frequency $\mathrm{Re}\,\Delta\omega = \omega-\Omega_0$, the background gain-loss $\mathrm{Im}\,\Delta\omega=\gamma$, or the quadratures of the arm-propagation coefficients $\kappa_{1}$ and $\kappa_{2}$, 
and facilitate
the complete reconstruction of all system and sensing parameters (see Appendix~\ref{app:paramrec}).

The topological significance of these working points is revealed 
when one studies their parametric dependence 
for fixed $c=|\kappa_2/\kappa_1|$
as the phase difference $\Delta\phi=\mathrm{arg}\,\kappa_2/\kappa_1$ changes from 0 to $2\pi$.
This dependence is displayed by the braiding diagrams 
in Fig.~\ref{fig:topological_evolution}(a,d).
Furthermore, when these trajectories are projected into the complex plane, the corresponding braiding phases afford a second topological interpretation
in terms of line-gap and point-gap phases \cite{Kawabata2019}, as described in Fig.~\ref{fig:topological_evolution}(b,c).
The braiding and gap topology changes at the same critical values $c_\mathrm{crit}^{d,o}$, Eq.~\eqref{eq:crit}, as observed for the winding topology of the bright interferometer signal, with an additional transition occurring in the off-diagonal configuration at the multimode EP condition $c=1\equiv c^\mathrm{MEP}$.
At this critical value, both source-detector configurations exhibit simultaneous dark-fringe operation at $\omega=\Omega_0, \gamma=\Gamma$.
Collecting these topological phase transitions, we therefore identify three critical working points (CWPs), 
given by
\begin{align}
&\omega=\Omega_0, \,\,\gamma=0,\,\,  \kappa_2/\kappa_1=c^{d}_\mathrm{crit}\,\,&&\mbox{(diag. CWP)},
\\
&\omega=\Omega_0, \,\, \gamma=0,\,\,  \kappa_2/\kappa_1=c^{o}_\mathrm{crit}\,\,&&\mbox{(offd. CWP)},
\\
&\omega=\Omega_0, \,\, \gamma=\Gamma,\,\,  \kappa_2/\kappa_1=c^\mathrm{MEP}\,\,&&\mbox{(MEP CWP)}.
\label{eq:CWP}
\end{align}
By setting the arm-propagation coefficient $\kappa_1$ to a suitable reference value, these CWPs can be enforced at any desired value of the counterpart coefficient $\kappa_2$.

\begin{figure*}
    \centering
    \includegraphics[width=0.85\linewidth]{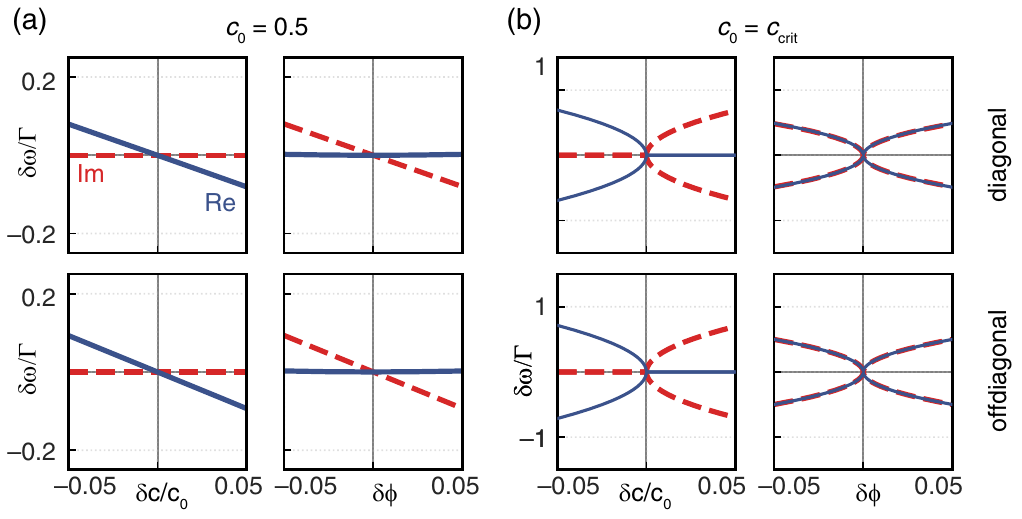}
    \caption{Metrological decoupling and enhanced sensitivity of dark-fringe working points. Complex spectral shifts $\delta\omega/\Gamma$ of the diagonal (top row) and off-diagonal (bottom row) dark-fringe working points in response to relative amplitude ($\delta c/c_0$) and phase ($\delta\phi$) perturbations of the arm-propagation coefficient ratio $\kappa_2/\kappa_1=ce^{i\Delta\phi}$ around a working point $\kappa_2/\kappa_1=c_0$. (a) In the linear operating regime away from the critical points (here illustrated for $c_0=0.5$), the sensing response is perfectly decoupled. Relative amplitude perturbations strictly induce real frequency shifts (solid blue), while phase perturbations strictly induce imaginary shifts (dashed red), establishing orthogonal metrological axes in the complex frequency plane. (b) At the critical working points ($c_0=c_{\mathrm{crit}}$), this linear separation breaks down. The dark-fringe states coalesce, and the exceptional interferometer exhibits a nonanalytical square-root dependence ($\delta\omega \propto \sqrt{\delta c/c_0 + i\delta\phi}$). 
    This behavior is characterized by pitchfork bifurcations and symmetric cusps in the complex frequency plane. In all panels, $J=1$ and $\Gamma=0.1$.
}
    \label{fig:perturbation}
\end{figure*}

\begin{figure*}
    \centering
    \includegraphics[width=0.85\linewidth]{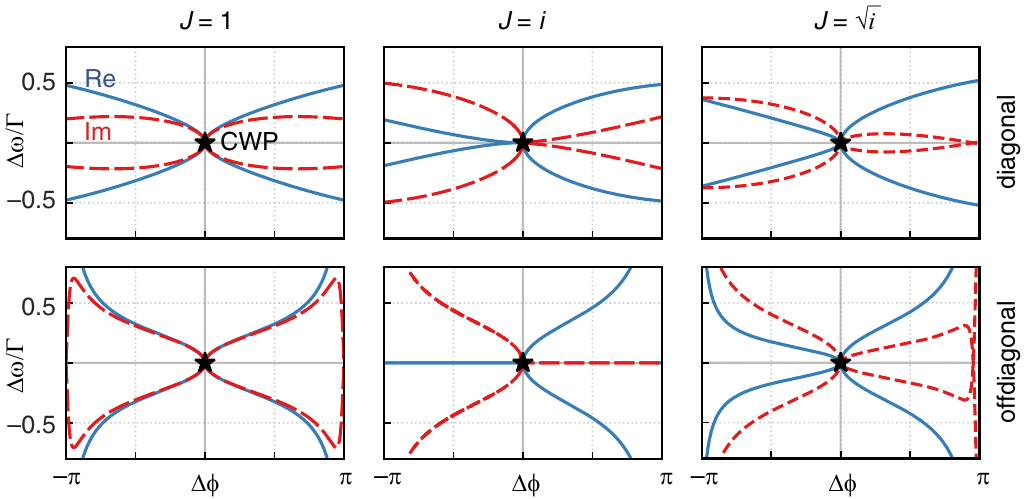}
    \caption{Control of complex frequency branch cuts and reconfigurable sensing axes. Real (blue) and imaginary (dashed red) frequency branches of the diagonal (top row) and off-diagonal (bottom row) dark-fringe detuning frequencies $\Delta\omega$ across the optical phase difference $\Delta\phi \in [-\pi, \pi]$ for $c=c_\mathrm{crit}$. The black star denotes the critical working point at $\Delta\phi=0$, where the roots coalesce and the topological phase transition occurs. From left to right, the mutual coupling is tuned from purely conservative ($J=1$) to purely dissipative ($J=i$), and to a complex intermediate value ($J=\sqrt{i}$). By engineering the complex phase of this internal coupling parameter $J$, the orientation of the nonanalytical square-root branch cuts and the resulting pitchfork bifurcations are deterministically rotated in the complex frequency plane. This confirms that the metrological axes of the exceptional interferometer can be fully reconfigured without altering the intrinsic spectral degeneracy of the collective multimode exceptional point. In all panels, $\Gamma=0.1$.
}
    \label{fig:complex_JCWP}
\end{figure*}

\subsection{Critical dark-fringe operation}
To elucidate the metrological impact of these CWPs, we 
configure the exceptional interferometer as a multiparameter sensor, capable of simultaneously resolving changes in both quadratures of these complex arm-propagation coefficients. For this, we map perturbations 
\begin{equation}
\kappa_2/\kappa_1=c_0e^{i\Delta \phi_0}(1+\delta c/c_0+i\delta\phi),     
\end{equation}
where $\delta c$ and $\delta \phi$ represent amplitude and phase deviations from reference values $c_0$ and $\Delta\phi_0$,
to complex spectral shifts $\delta\omega$ of the dark-fringe working points \eqref{eq:wp_modes}.
Away from the 
CWPs, a Taylor expansion yields the linear shifts 
\begin{equation}
\begin{aligned}
\delta\omega_d^\pm &= \pm \frac{\Gamma J}{2\Delta\omega_d^\pm}\sqrt{c_0e^{i\Delta\phi}}(\delta c/c_0+i\delta\phi)&&\mbox{(diagonal)}
\\
\delta\omega_o &= -\frac{2\Gamma J}{\Delta\omega_o}\frac{c_0e^{i\Delta \phi_0}}{(1+c_0 e^{i\Delta \phi_0})^2}(\delta c/c_0+i\delta\phi)
&&\mbox{(off-diagonal)}
,
\end{aligned}
\end{equation}
cleanly separating the two quadrature changes into orthogonal directions in the complex frequency plane, detected as shifts of the dark-fringe frequency $\omega$ and the background loss parameter $\gamma$.
As elaborated further below, the orientation of these metrological axes in the sensing parameter space can be controlled by the choice of the resonator parameters.

A key feature of these frequency shifts 
is their inverse proportionality in the working-point detuning frequency $\Delta\omega_{o,d}$, resulting in a strongly enhanced parametric response as the 
diagonal and off-diagonal CWPs are approached. 
This
singular behavior
reflects the crossing of the resonance condition $\Delta\omega=0$ with the dark-fringe condition $S_i=0$ near the multimode EP $c=1=c^\mathrm{MEP}$, where the leading-order bright-fringe signal becomes suppressed, resulting in exact signal synchronization [see upper inset in Fig.~\ref{fig:main_model}(e)]. 
At these CWPs, the dark-fringe frequency shifts fundamentally change their order from a linear to a nonanalytical square-root dependence $\delta\omega \propto \sqrt{\delta c/c_0 + i\delta\phi}$, with
\begin{equation}
\begin{aligned}
(\delta\omega_d^\pm)^2 &= -\Gamma J(1-\Gamma/2J)(\delta c/c_0+i\delta\phi),\\
(\delta\omega_o)^2 &= -\Gamma J(1-\Gamma^2/4J^2)(\delta c/c_0+i\delta\phi)
.
\end{aligned}
\label{eq:wp_critical_shifts}
\end{equation}
This corresponds to a critical  dependence of the detuning frequency on the amplitude and phase deviations as the interferometer operates at the topological phase transitions separating the different bright-fringe and dark-fringe operating regimes, as
illustrated in the lower inset in Fig.~\ref{fig:main_model}(e) for $\Gamma/J=0.1$ with real $J$.

The detailed evaluation of the real and imaginary frequency branches is shown in Figs.~\ref{fig:perturbation} and \ref{fig:complex_JCWP},
displaying the metrological response to relative amplitude ($\delta c/c_0$) and phase ($\delta\phi$) perturbations of the arm-propagation coefficient ratio $\kappa_2/\kappa_1=ce^{i\Delta\phi}$ around a working point $\kappa_2/\kappa_1=c_0$.
Away from the critical thresholds, as illustrated in Fig.~\ref{fig:perturbation}(a) for $c_0=0.5$, the sensing response to both perturbations is decoupled. In this linear operating regime, relative amplitude perturbations  induce real frequency shifts, while phase perturbations  induce imaginary shifts, establishing orthogonal metrological axes in the complex frequency plane. However, as shown in Fig.~\ref{fig:perturbation}(b), when the system is tuned to the CWP ($c_0=c_{\mathrm{crit}}$), where the two dark-fringe working points coalesce at $\Delta\omega=0$, this linear separation is replaced by the nonanalytical square-root dependence established in Eq.~\eqref{eq:wp_critical_shifts}.
The versatility of this sensing mechanism is further expanded by controlling the complex branch cuts that define these nonanalytical responses, as shown in Fig.~\ref{fig:complex_JCWP}. By engineering the complex phase of the mutual coupling parameter $J$, the system can be transitioned from a purely conservative ($J=1$) to a purely dissipative ($J=i$) 
coupling regime,
as well as to intermediate complex couplings ($J=\sqrt{i}$).
This parameter tuning rotates the orientation of the nonanalytical square-root branch cuts in the complex frequency plane. Consequently, the metrological axes of the exceptional interferometer can be reconfigured to target specific sensing modalities,
leading to a versatile interferometry principle that realizes critical non-Hermitian sensing directly under dark-fringe operating conditions.

\section{Conclusions and outlook \label{sec:outlook}}
In summary, we have developed a non-Hermitian interferometry principle in which dark-fringe frequencies display a nonanalytical dependence on the signal-propagation coefficients through the interferometer arms. In a concrete minimal model, this behavior is induced by replacing the conventional beam splitters of canonical Michelson or Mach-Zehnder interferometers with spectrally defective non-Hermitian two-mode resonators. While these resonant beam splitters display single-mode EPs, 
the interferometer arms act as a tunable external control that induces a multimode EP response.
The resulting exceptional interferometry framework offers a rich variety of operating modes that are organized by non-Hermitian symmetry and topology. 
In the bright-fringe regime,  the multimode EP response results in a strong signal suppression, 
stemming from a transition between quartic and cubic super-Lorentzian lineshapes that occurs without altering the intrinsic resonance frequency of the system.
The nonanalytical dark-fringe operation then emerges at critical working points, determined by
topological winding and braiding phase transitions that separate the different
bright-fringe regimes.
This parameter dependence replicates the sensitivity enhancement of resonant EP sensors, but here is attained under interferometric dark-fringe conditions. 
Consequently, the critical working points strongly enhance the dark-fringe frequency response to small perturbations, reflecting critical scaling at the topological phase transitions.
This establishes a high-precision sensing scenario that directly exploits the multimode character of the EP.

We have demonstrated this general concept using a concrete non-Hermitian resonator model based on standard components that are readily available across a wide range of physical platforms. The required non-Hermitian beam-splitting elements can be implemented, for example, through coupled waveguides with engineered gain-loss contrast \cite{Ruter2010,QuirozJuarez:19,On2024} and microring resonators supporting asymmetric coupling pathways \cite{Van2016Microring,microjan2,
Zhong2020,Zhong2019,Soleymani2022,Kullig:23}. These platforms naturally accommodate the complex-valued mode interactions necessary to realize EPs in the resonant response, and they readily serve the same purpose in dark-fringe operation, thereby enabling interferometric responses that transcend the limits of conventional designs.
Consequently, a broad class of existing non-Hermitian beam-splitting setups can be configured to realize the proposed exceptional interferometer principle. As this principle is governed by universal non-Hermitian symmetries and robust topology, it is agnostic to the specific photonic implementation, and can readily benefit from future advancements in non-Hermitian beam-splitter technology. Furthermore, the underlying framework is highly transferable, making the proposed  principle directly applicable to other non-Hermitian settings, including microwave, optomechanical, electronic, and acoustic systems
\cite{Dembowski2001,Xu2016,Schindler2012,zhu2014pt}.

The present work focuses on the classical interferometer signal relevant to most practical applications. A natural direction for future research is the exploration of the quantum limit, which is critical for applications in fundamental precision interferometry. In conventional resonant EP devices, quantum-noise amplification counteracts the enhanced parametric sensitivity \cite{Yoo2011,Wiersig:20,Langbein2018}, and analogous constraints are anticipated for exceptional interferometry under bright-fringe operating conditions. However, since in dark-fringe operation the system primarily serves as an absorber rather than an emitter, it is expected to be  less susceptible to such additional noise sources \cite{Caves1981}. Beyond 
chip-scale quantum devices such as superconducting  circuits \cite{Naghiloo2019}, it would be particularly interesting to explore the integration of exceptional dark-fringe sensing into macroscopic optomechanical interferometers, such as those utilized for gravitational-wave detection \cite{ligo}. 
With its wide range of prospective implementations and applications, the proposed exceptional interferometry paradigm serves as a highly promising addition to the available tools in precision metrology,
establishing a general pathway to novel sensing functionalities through a multimode-enhanced non-Hermitian topological interference landscape.

The data that support the findings of this article and the code used to analyze the data are publicly available \cite{researchdata}.
\begin{acknowledgments}
This research was funded by EPSRC via Grant No. EP/W524438/1.
\end{acknowledgments}

\appendix

\section{Input-output description of the non-Hermitian resonant beam splitters\label{app:nhresonator}}

\begin{figure*}[ht]
    \centering
    \includegraphics[width=0.7\linewidth]{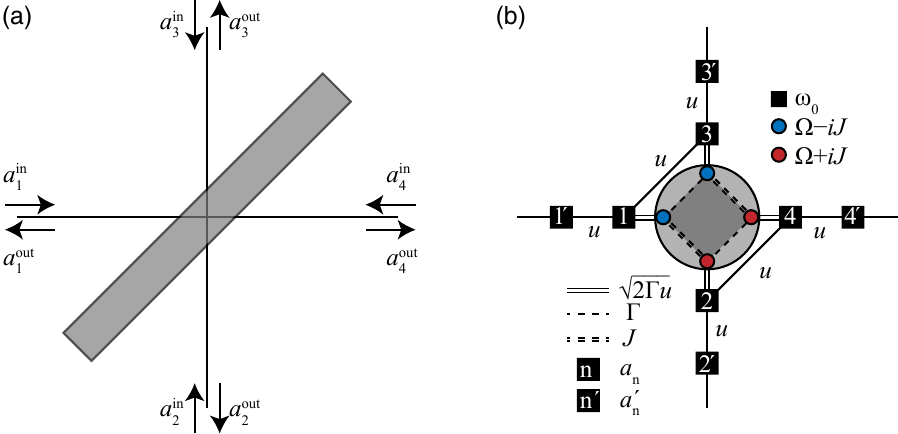}
    \caption{(a) Definition of incoming and outgoing modes for a beam splitter. (b) Implementation of the non-Hermitian resonant-cavity beam splitter within coupled-mode theory.}
    \label{fig:inoutEX}
\end{figure*}

In the main text we utilize an effective two-level Hamiltonian describing the transmission through the non-Hermitian resonant-cavity beam splitter. We arrive at this effective Hamiltonian from a standard coupled-mode perspective  \cite{Fan:03}, relying on established coupling configurations \cite{Zhong2020}, where the optical resonator is treated as a localized subsystem coupled to continuum modes representing propagating fields in the external waveguides, see Fig.~\ref{fig:inoutEX}.

A beam splitter involves four input and four output channels, with amplitudes denoted by $a_{n}^\mathrm{in,out}$, $n=1,2,3,4$, as shown in Fig.~\ref{fig:inoutEX}(a). The channels imprint amplitudes
\begin{align}
    a'_n&=a_n^\mathrm{in}+a_n^\mathrm{out}, \quad 
    a_n=-ia_n^\mathrm{in}+i a_n^\mathrm{out}
    \label{eq:propmodesMETH}
\end{align}
onto localized coupled modes, labeled according to the positions indicated in Fig.~\ref{fig:inoutEX}(b). 
This adopts the wide-band limit,  which removes the dispersion in the continuum modes, corresponding to a band dispersion $\omega=\omega_0+2u\cos(k)$ with frequency offset $\omega_0$ and bandwidth parameter $u\gg |\omega-\omega_0|$, and operating near the band center with dimensionless wave number $k=\pm \pi/2$.

We configure the non-Hermitian  beam splitter as a cavity with
four internal real-space modes, whose amplitudes we denote by $b_{1,2,3,4}$. As our main design choice, we specify the bare internal Hamiltonian of this cavity as
\begin{align}
    H_0&=\begin{pmatrix}
        \Omega-iJ  & J & \Gamma & 0\\
        J & \Omega+iJ & 0 & \Gamma \\
        \Gamma & 0 & \Omega-iJ & J\\
        0 & \Gamma & J & \Omega+iJ          
    \end{pmatrix}
  \nonumber \\ & 
    =\Omega \openone+J(X_2-iZ_2)+\Gamma X_1,
\end{align}
with the coupling $J$ and bare resonance frequency $\Omega$. For compactness, we introduced the matrices $X_1= \sigma_x\otimes \openone$, $X_2= \openone\otimes \sigma_x$, $Z_2=\openone\otimes \sigma_z$ with standard Pauli matrices $\sigma_r$.
We confirm reciprocity via the relation $H=H^T$, and PT symmetry 
\cite{ElGanainy2018}
via the relation $(H-\Omega)^*=P(H-\Omega)P^{-1}$, where $P=X_2=P^{-1}$ represents the  parity operator, and all coupling coefficients are taken real.

The external channels couple to the internal modes with strength
$W=\sqrt{2\Gamma u}\openone$,
where the scaled coupling strength $\Gamma$ is assumed to be frequency-independent over the relevant spectral window. 
In the interferometric context, any additional dispersion in the continuum modes can be taken into account in the frequency-dependence of the cumulative arm-propagation coefficients $\kappa_1$ and $\kappa_2$.
In vector notation, 
the coupled-mode equations are then given by
\begin{align}
&\omega\mathbf{a}
    =u\mathbf{a}'
    +(\omega_0+uX_1)  \mathbf{a}
    +\sqrt{2\Gamma u}
\mathbf{b}    ,
    \nonumber
    \\
    &
    \omega\mathbf{b} 
    =H_0\mathbf{b} +\sqrt{2\Gamma u}
        \mathbf{a} .
\end{align}

We use the second equation to eliminate $\mathbf{b}$, take the wide-band limit  $u\gg |\omega-\omega_0|$, and pass over to the 
propagating modes via equation \eqref{eq:propmodesMETH}, giving
  $ \mathbf{a}^\mathrm{out} =S\mathbf{a}^\mathrm{in}$
with scattering matrix 
\begin{align}
    S=\openone-2[\openone+iX_1+2i\Gamma(\omega-H_0)^{-1}]^{-1}.
\end{align}
Written explicitly, this scattering matrix takes the block form
\begin{align}
 S=   \begin{pmatrix}
        0 &  t_2\\ t_1 & 0 
    \end{pmatrix},
\end{align}
where the transmission blocks 
$t_1=t_2=i\openone+2\Gamma[(\omega+i\Gamma)\openone -H_\mathrm{eff}]^{-1}$ with the effective two-level Hamiltonian $H_\mathrm{eff}$ stated in the main text, completing the construction. Analyzing the pole structure of the scattering matrix, we verify that the system is dynamically stable as long as $\gamma+\Gamma\geq 0$, supporting the feasibility of the system using standard optical components.

\section{Multimode exceptional point analysis\label{app:mmep}}

As described in the main text, the total transmission matrix 
\eqref{eq:main_result} can be rewritten in terms of a nilpotent effective Hamiltonian $N$, see Eqs.\eqref{eq:teff}-\eqref{eq:neff}. The singular change of the lineshape from a fourth-order super-Lorentzian to a third-order super-Lorentzian is then obtained from the expansion \eqref{eq:teff2}, whose structure is dictated by the nilpotency of $N$, which we here examine in detail.

For a compact discussion, we write 
\begin{align}
N=    \Gamma(X\otimes Z-iZ\otimes\openone)-iJ\openone\otimes Z
\nonumber\\+J\kappa_+Z\otimes X-iJ\kappa_-X\otimes X
\end{align}
with $X=\sigma_x$, $Y=\sigma_y$, $Z=\sigma_z$ given by the standard Pauli matrices, and $\kappa_{\pm}=(\kappa_1\pm \kappa_2)/(2\sqrt{\kappa_1\kappa_2})$ as defined in the main text. 
We then evaluate
\begin{align}
N^2=&2 \Gamma J[
-iX\otimes \openone-Z\otimes Z
+
\kappa_+ (Y\otimes Y-i\openone\otimes X)],
 \\
 N^3=&4\Gamma J^2\kappa_-[
-\openone\otimes X
-iY\otimes Y
\nonumber \\ &
+\kappa_+(iZ\otimes Z
-X\otimes \openone)
+
 \kappa_-(X\otimes Z
-iZ\otimes \openone)],
\\
N^4=&0.
\end{align}
For $\kappa_1=\kappa_2$, we have $\kappa_-=0$, hence $N^3= 0$, as stated in the main text.

To determine the exact multimode character of the EP, 
we inspect the corresponding Jordan normal form, amounting to a  similarity transformation $J=TNT^{-1}$ where the possible options are
\begin{align}
    J&=\left(\begin{array}{cccc}
        0 & 1 & 0 & 0 \\
        0 & 0 & 1 & 0 \\
        0 & 0 & 0 & 1 \\
        0 & 0 & 0 & 0 \\        
    \end{array}\right)
    \qquad\mbox{(single-mode EP$_4$)},
\\
J&=\left(\begin{array}{ccc|c}
        0 & 1 & 0 & 0 \\
        0 & 0 & 1 & 0 \\
        0 & 0 & 0 & 0 \\
        \hline
        0 & 0 & 0 & 0 \\        
    \end{array}\right)
    \qquad\mbox{(two-mode EP$_{3,1}$)},
\\
J&=\left(\begin{array}{cc|cc}
        0 & 1 & 0 & 0 \\
        0 & 0 & 0 & 0 \\ \hline
        0 & 0 & 0 & 1 \\
        0 & 0 & 0 & 0 \\        
    \end{array}\right)
    \qquad\mbox{(two-mode EP$_{2,2}$)},
\\
J&=\left(\begin{array}{cc|c|c}
        0 & 1 & 0 & 0 \\
        0 & 0 & 0 & 0 \\ \hline
        0 & 0 & 0 & 0 \\ \hline
        0 & 0 & 0 & 0 \\        
    \end{array}\right)
    \qquad\mbox{(three-mode EP$_{2,1,1}$)}.
\end{align}
The multimode scenarios are characterized by a fragmentation of the  Jordan normal form into smaller blocks, indicated by the lines, where each block is associated with a distinct eigenmode.  The block sizes determine the partial multiplicity of the eigenmodes, which we indicate by the indices. 

We side-step the explicit construction of the similarity transformation $T$ and determine the basis-invariant ranks 
\begin{align}
    \rnk J^{(1,2,3)}&= (3,2,1)
    \qquad\mbox{(single-mode EP$_4$)},
\\
    \rnk J^{(1,2,3)}&= (2,1,0)
    \qquad\mbox{(two-mode EP$_{3,1}$)},
\\
    \rnk J^{(1,2,3)}&= (2,0,0)
    \qquad\mbox{(two-mode EP$_{2,2}$)},
\\
    \rnk J^{(1,2,3)}&= (1,0,0)
    \qquad\mbox{(three-mode EP$_{2,1,1}$)},
\end{align}
directly from the derived matrices $N^l$ in their original basis \cite{bid2024uniform}, amounting to counting the number of finite eigenvalues of $(N^l)^\dagger N^l$. For $\kappa_1\neq\kappa_2$, we obtain 
\begin{equation}
\rnk N=3, \qquad \rnk N^2=2, \qquad
\rnk N^3=1,
\end{equation}
which corresponds to an ordinary single-mode EP$_4$.
For $\kappa_1=\kappa_2$, we instead find 
\begin{equation}
\rnk N=2, \qquad \rnk N^2=1, \qquad
\rnk N^3=0.
\end{equation}
This determines that the exceptional interferometer realizes a two-mode EP$_{3,1}$.

\section{Pole-compensated topological winding invariant\label{app:polecomp}} 
To characterize the discrete operational dark-fringe phases of the non-Hermitian interferometer, we relate the number $\mathcal{Z}$ of dark-fringe  working points at a given set of system parameters to the topological invariant $\mathcal{W}$ tracking the frequency-dependent winding of the signal amplitudes $t_{11}(\omega)/\kappa_1$ and $t_{12}(\omega)/\kappa_1$ in the complex plane. We define $\psi_r=\mathrm{arg}\,t_{r1}(\omega)/\kappa_1$ as the continuously unwound, accumulated, signal phase in each detector configuration, $r=1,2$.
The winding of this phase can be split into a contribution from the quadruple pole $\propto(\Delta \omega+i\Gamma)^4$ in the denominator and $\mathcal{Z}$ dark-fringe zeros in the numerator of the analytical expressions,  obtained from the diagonal or off-diagonal elements of the total transmission matrix $t(\omega)$, Eq.~\eqref{eq:main_result}.
As each pole gives a signal-phase winding contribution of $+\pi$ and each zero  a signal-phase winding  contribution of $-\pi$, 
\begin{equation}
    \label{eq:winding_def}
    \mathcal{W} = \frac{1}{2\pi} \ \oint_{C} \partial_{\omega} \psi_{r} \, \mathrm{d}\omega  =\frac{4\pi-\mathcal{Z}\pi}{2\pi},
\end{equation}
hence, $\mathcal{Z}=4-2\mathcal{W}$.
In the definition above, the integration contour $C$ is shifted into the complex plane via the analytic continuation $\omega \to \omega + i\varepsilon$ with $\varepsilon>0$, physically corresponding to addition of small background losses (see also  Fig.~\ref{fig:finite_loss}).
The analytical continuation with 
 $\varepsilon<0$ changes the winding number to
 $\widetilde{\mathcal{W}}=\frac{4\pi+\mathcal{Z}\pi}{2\pi}$, so that then 
$\mathcal{Z}=2\widetilde{\mathcal{W}}-4$.

\section{Braid group classification of dark-fringe operating regimes\label{app:braiding}}
In non-Hermitian systems, the configuration of complex singularities in a multi-dimensional parameter space can be classified using Artin braid groups $\mathbf{B_n}$, where $\mathbf{n}$ represents the number of discrete states. The braids describe the topology of strands formed when changing a parameter.
The braid group can be flattened into the permutation group $\mathbf{S_n}$, which describes the resulting permutation of the singularities. 
While existing literature has applied braid theory to the eigenvalues of non-Hermitian Hamiltonians, we demonstrate here how this topological framework applies to the detuning frequencies in the dark-fringe operating regime in the two source-detector configurations, given by Eq.~\eqref{eq:wp_modes} in the main text.

\begin{table*}[t]
\centering
\caption{Braid group classification of topological dark-fringe operating regimes. 
The braid words represent the exchange of the dark-fringe detuning frequency over a single fundamental $2\pi$ period of the phase difference $\Delta\phi$ accumulated in the interferometer arms.
}
\vspace{0.2em}
\label{tab:braid_topology}
\begin{tabular}{@{} l c c c c l @{}}
\toprule
\textbf{Readout Port} & \textbf{Asymmetry ($c$)} & \textbf{Gap Topology} & \textbf{Group} & \textbf{Period to Close} & \textbf{Braid Word ($2\pi$ sweep)} \\ 
\midrule
\multirow{3}{*}{\textbf{Diagonal ($t_{11}$)}} 
& $c < c^{d}_\mathrm{crit}$ & Imaginary Line-Gap & $\mathbf{B_4}$ & $4\pi$ & $\sigma_1\sigma_3$ (Inner/Outer Exchange) \\
& $c = c^{d}_\mathrm{crit}$ & Dark fringe at $\Delta\omega_d=0$ & - & - & Singularity (Gap Closing) \\
& $c > c^{d}_\mathrm{crit}$ & Nested Point-Gap & $\mathbf{B_4}$ & $8\pi$ & $\sigma_3 \sigma_2 \sigma_1$ (Coupled Exchange) \\ 
\midrule
\multirow{4}{*}{\textbf{Off-diagonal ($t_{21}$)}} 
& $c < c^{o}_\mathrm{crit}$ & Imaginary Line-Gap & $\mathbf{B_2}$ & $2\pi$ & $e$ (Identity) \\
& $c = c^{o}_\mathrm{crit}$ & Dark fringe at $\Delta\omega_o=0$ & - & - & Singularity (Gap Closing) \\
& $c^{o}_\mathrm{crit} < c < c^\mathrm{MEP}$ & Point-Gap & $\mathbf{B_2}$ & $4\pi$ & $\sigma_1$ \\
& $c = c^\mathrm{MEP}$ & Bright fringe at $\Delta\omega_o=0$ & - & - & Singularity (Gap Closing at $\infty$) \\
& $c > c^\mathrm{MEP}$ & Real Line-Gap & $\mathbf{B_2}$ & $2\pi$ & $e$ (Identity) \\ 
\bottomrule
\end{tabular}
\end{table*}

To identify this braiding topology, we consider the dependence of the detuning frequencies as a function of the phase difference $\Delta\phi$ in the arm-propagation  ratio $\kappa_2/\kappa_1=c e^{i\Delta\phi}$  while keeping its magnitude $c$ fixed. As the phase difference is continuously swept from $0$ to $2\pi$, the dark-fringe detuning frequencies in a given source-detector configurations weave through the three-dimensional space $(\mathrm{Re}[\Delta\omega_{d/o}], \mathrm{Im}[\Delta\omega_{d/o}], \Delta\phi)$ (see Fig.~\ref{fig:topological_evolution} in the main text). The resulting strands form topological knots, where the permutation of the dark state is governed by the braid generators $\sigma_i$, representing the counter-clockwise exchange of the $i$-th and $(i+1)$-th adjacent strands. 

In the off-diagonal  channel, the destructive interference condition generates two dark-fringe detuning frequencies $\pm\Delta\omega_o$, restricting the topology to the two-strand braid group $\mathbf{B_2}$. We then identify three regimes as depicted in Fig.~\ref{fig:topological_evolution}(c) in the main text.

\emph{Imaginary Line-Gap Regime} ($c < c^{o}_\mathrm{crit}$): In this regime, the dark-fringe detuning frequencies are real when $\Delta\phi\equiv 0 \pmod\pi$, and are confined to their respective halves of the complex plane, resulting in a  line gap along the imaginary axis. A $2\pi$ sweep returns each detuning frequency to its original position without exchange, corresponding to the trivial permutation. As the strands do not cross, this represents the trivial identity braid group element $e$.

\emph{Point-Gap Regime} ($c^{o}_\mathrm{crit} < c < c^\mathrm{MEP}$): The trajectories enclose the origin, forcing the evaluation of the square root across its principal branch cut. After a $2\pi$ sweep, the positive branch analytically continues into the negative branch, $\Delta\omega_{o\,+}(2\pi) = \Delta\omega_{o\,-}(0)$, so that both frequencies exchange their positions. This is the fundamental braid generator $\sigma_1$. 
    To  return the detuning frequencies to their original positions, the phase difference must be swept to $4\pi$.    
    
    \emph{Real Line-Gap Regime} ($c > c^\mathrm{MEP}$): As we further increase the magnitude of the 
arm-propagation coefficient ratio beyond $c=1$, the two trajectory branches join at infinity and then form again $2\pi$ periodic loops in the complex plane, which now are separated by a line gap along the real axis. At $\Delta\phi\equiv 0 \pmod\pi$, the dark-fringe detuning frequencies are now imaginary.  This again corresponds the identity braid group element $e$.

The diagonal channel features four distinct interferometric dark-fringes (two inner, two outer), elevating the system to the four-strand braid group $\mathbf{B_4}$ with generators $\sigma_1, \sigma_2, \sigma_3$. The analytical expression of $\Delta\omega_d$ (see Eq.~\eqref{eq:wp_modes} in the main text) includes an additional square root, which results in nested Riemann sheets. We then identify two regimes as illustrated in Fig.~\ref{fig:topological_evolution}(b) in the main text.

\emph{Imaginary Line-Gap Regime} ($c < c^{d}_\mathrm{crit}$): The additional square root enforces an exchange between the inner and outer root manifolds, so that two  pairs of strands wind around each other, corresponding to the braid element $\sigma_1\sigma_3$. The detuning frequencies only return to their positions 
   after the phase is advanced by $4\pi$.

\emph{Nested Point-Gap Regime} ($c > c^{d}_\mathrm{crit}$): When the system surpasses the critical point, the trajectory branches join up into a single loop with point-gap topology. A 2$\pi$ sweep of the phase difference now forces the dark-fringe detuning frequencies to exchange identity cyclically, so that they return to their original positions only after a $8\pi$ sweep. For a $2\pi$ sweep, the corresponding braid is governed by the composite word $\sigma_3 \sigma_2 \sigma_1$.

The described dark-fringe braiding topology is summarized in Table \ref{tab:braid_topology}. 

\section{Parameter reconstruction from dark-fringe working points\label{app:paramrec}}

The operating characteristics of the exceptional interferometer depend on the non-Hermitian beam splitter parameters $J$, $\Gamma$, and $\Omega=\Omega_0-i\gamma$, where the latter enters the frequency detuning parameter $\Delta\omega=\omega-\Omega$.
The frequency-dependent signals in the two source-detector configurations then provide interferometric sensing of the effective arm-propagation coefficients $\kappa_1$ and $\kappa_2$.
To verify the possible extraction of these system and sensing parameters, we describe here procedures based on the dark-fringe working points \eqref{eq:wp_modes}. Let us first assume that the quadratures fulfill the equal-phase condition $\mathrm{Im}\,\kappa_1/\kappa_2=0$, so that the dark-fringe working points in both source-detector configurations occur in a fixed setup at real frequency detunings.
By further combining the values of these frequency detunings
into the expression
\begin{align}
2\frac{(\Delta\omega_d^\pm)^2-(\Delta\omega_o)^2}{
    (\Delta\omega_d^+)^2-(\Delta\omega_d^-)^2}
    =
    \frac{2\sqrt{\kappa_1\kappa_2}}{\kappa_1+\kappa_2}\pm 1,
    \label{eq:combined}
\end{align}
they determine the ratio of the geometric and arithmetic mean of the cumulative arm-propagation coefficients $\kappa_1$ and $\kappa_2$, 
independently of the internal interferometer parameters $\Gamma$ and $J$ characterizing the non-Hermitian resonator. 
As in conventional interferometry, information about these internal parameters can be obtained from suitable reference configurations, e.g., by determining the dark-fringe working points in a configuration with $\kappa_1=\kappa_2$ or $\kappa_2=0$. Furthermore, exploiting the specific characteristics of the exceptional interferometer, these parameters can be inferred in any given setup  from the additional relations
\begin{align}
(\Delta\omega_d^+)^2+(\Delta\omega_d^-)^2
    &=2\Gamma(2J-\Gamma),
    \nonumber
\\
(\Delta\omega_d^+)^2-(\Delta\omega_d^-)^2
    &=4\Gamma J\sqrt{\kappa_2/\kappa_1}.
\end{align}
Thereby, the dark-fringe working points deliver a full characterization of the interferometer and sensing parameters.

\section{Additional numerical modeling of the operating parameter space} 

Here, we provide additional numerical results
modeling the parameter space dictating bright and dark fringe interferometry.

\begin{figure*}[t]
    \centering
    \includegraphics[width=0.95\linewidth]{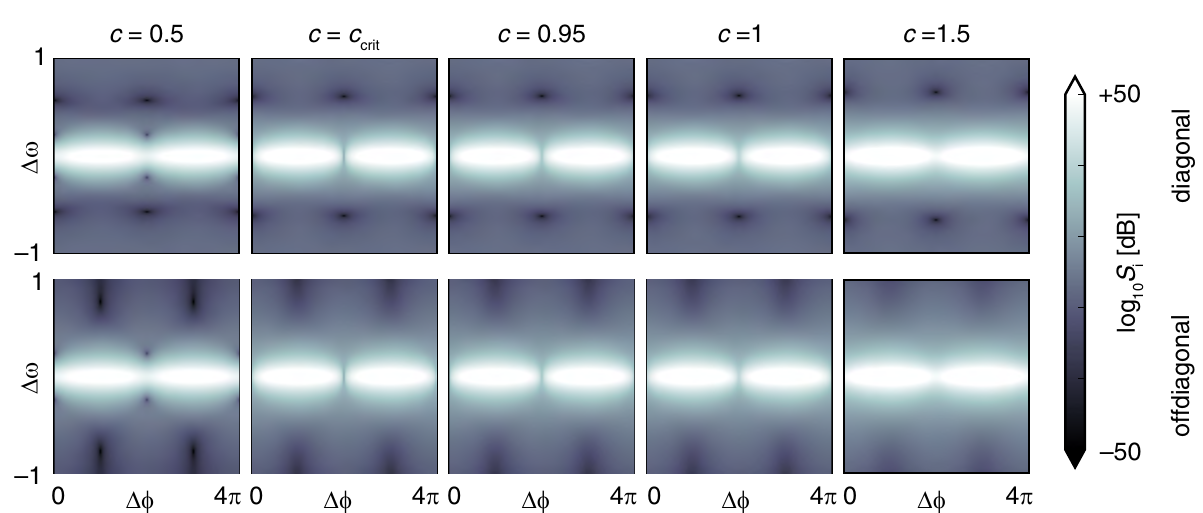}
    \caption{Frequency-resolved intensity against cumulative arm-propagation phase difference. The signal intensity $S_1=|t_{11}/\kappa_1|^2$ in the diagonal channel (top row) and $S_2=|t_{21}/\kappa_1|^2$ in the off-diagonal channel (bottom row) is shown as a function of frequency detuning $\Delta\omega=\omega-\Omega$ and two periods of the phase difference $\Delta\phi=\mathrm{arg}\,(\kappa_2/\kappa_1)$, with $c=|\kappa_2/\kappa_1|$ set to representative values $c= 0.5$, $c^{o/d}_{\mathrm{crit}}$, $0.95$, $1.0$, $1.5$ in different operating regimes.
    At the phase transition $c=c^{o/d}_{\mathrm{crit}}$, the dark-fringe detuning frequency collapses with the bright-fringe resonant condition $\Delta\omega=0$.
In all panels, $J=1$, $\Gamma=0.1$, $\gamma=0$.
    }
    \label{fig:fringefigure}
\end{figure*}

Fig.~\ref{fig:fringefigure} maps the frequency-resolved intensity landscape as a function of the phase difference $\Delta\phi$ across two full periods, allowing the identification of dark and bright fringes across representative values of the asymmetry parameter $c$. In agreement with the analytical framework derived in the main text, the dark-fringe working points intersect the bright-fringe resonance ($\Delta\omega=0$)  at the critical threshold $c=c^{o/d}_{\mathrm{crit}}$. Beyond this threshold, there are no dark fringes at the resonance condition $\Delta\omega=0$ in either the diagonal or off-diagonal source-detector configurations. However, away from the resonance, the dark fringes remain present across the entire parameter space for the diagonal configuration. This is a direct mathematical consequence of the nested radical structure 
of the dark-fringe detuning frequency $\Delta\omega_d$ (see Eq.~\eqref{eq:wp_modes} in the main text).

\begin{figure*}[t]
    \centering
    \includegraphics[width=\linewidth]{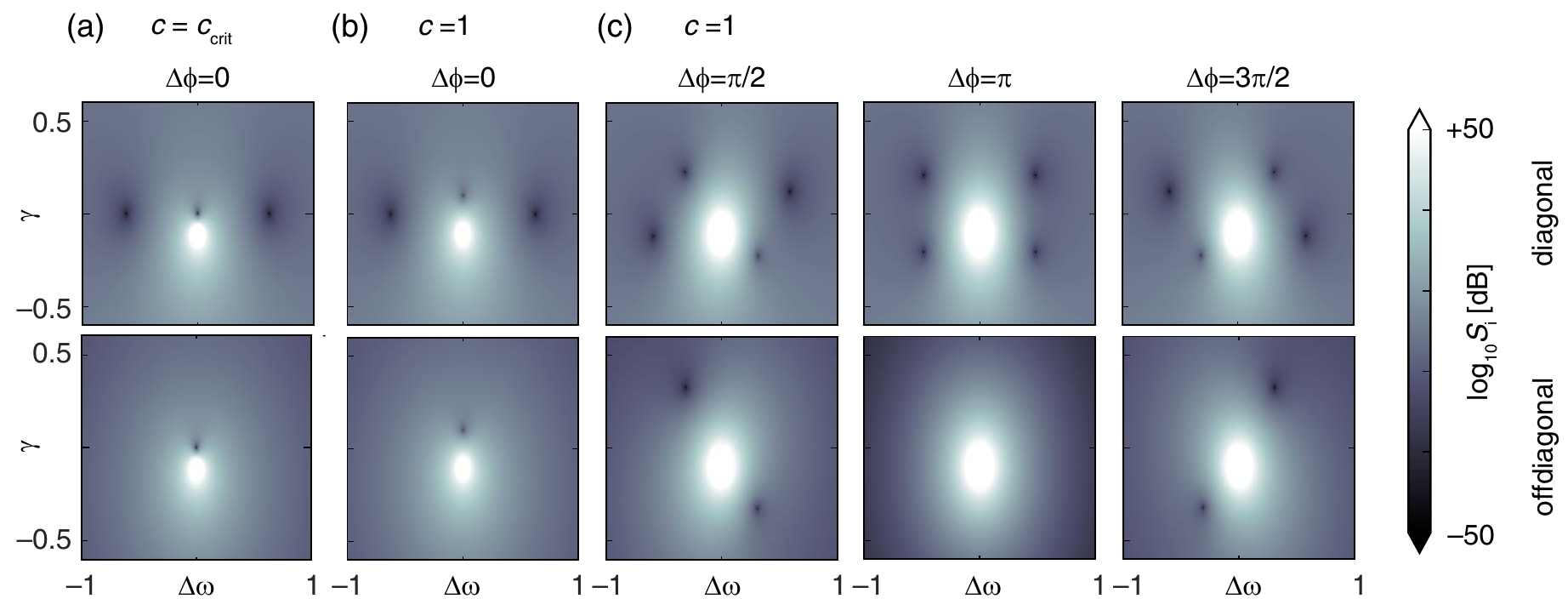}
\caption{Distinct critical working points in bright-dark mixed mode operating regimes. Logarithmic intensity maps of the diagonal ($|t_{11}/\kappa_1|^2$, top row) and off-diagonal ($|t_{21}/\kappa_1|^2$, bottom row) channels in the complex frequency plane,
spanned by the detuning frequency $\Delta\omega$  and the background loss parameter $\gamma$.
(a) At the critical values $c = c_{\mathrm{crit}}$ 
of the arm-propagation coefficient ratio $\kappa_2/\kappa_1=ce^{i\Delta\phi}$, the dark-fringe working point collapses with the EP resonance at $\Delta\omega=\gamma=0$. (b) shows corresponding intensity maps in the presence of the multimode EP, $c = c^{\mathrm{MEP}} = 1$ and $\Delta\phi = 0$, resulting in a \sout{simultaneous} dark-fringe working point in both channels at $\Delta\omega=0$, $\gamma=\Gamma$. (c) shows how these dark-fringe working points migrate around the multimode EP for finite phase differences $\Delta\phi$.
In all panels, $J=1$ and $\Gamma=0.1$.
}
    \label{fig:CWP}
\end{figure*}

\begin{figure*}[t]
    \centering
    \includegraphics[width=\linewidth]{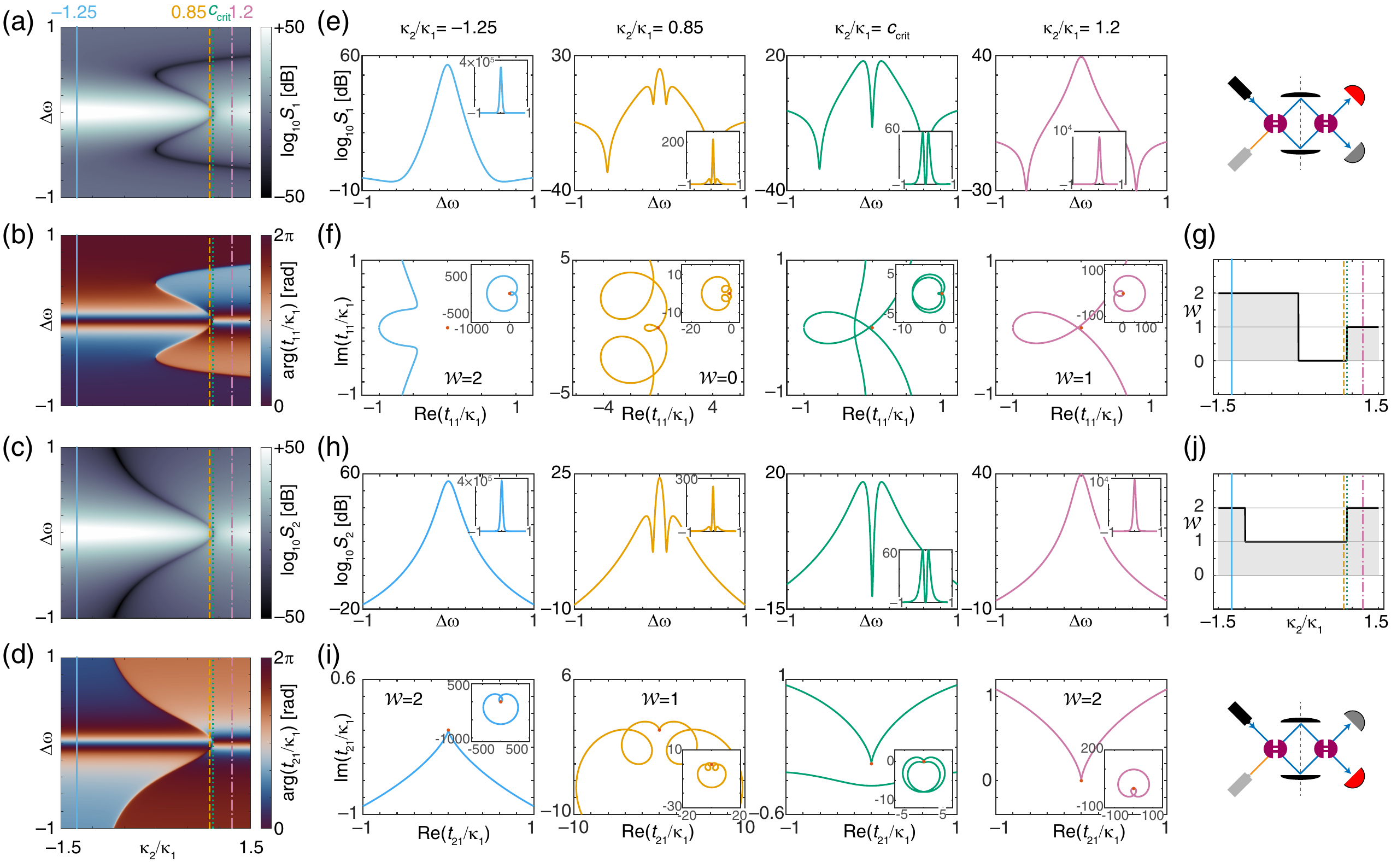}
    \caption{Effect of finite background loss. Analogous to Fig.~\ref{fig:main_results} in the main text, but for finite background loss $\gamma=0.01$. While the introduction of dissipation shifts the exact spectral coordinates of the bright and dark fringes, the overall non-Hermitian topology remains intact. The Argand trajectories and their corresponding quantized winding numbers $\mathcal{W}$ continue to exhibit distinct discrete plateaus. Consequently, the abrupt topological phase transitions dictating the emergence and annihilation of dark-fringe pairs, now given by Eq.~\eqref{eq:critsup}, survive intact, confirming the structural robustness of the exceptional interferometer against experimental imperfections.
}
    \label{fig:finite_loss}
\end{figure*}

To further examine the origins of these distinct operating regimes, Fig.~\ref{fig:CWP} maps the intensity landscape in the complex frequency plane, spanned by the real detuning $\Delta\omega$ and the imaginary background loss parameter $\gamma$. 
This representation isolates the critical working points (CWPs) introduced in the main analysis of operating regimes as exact intensity zeros.
Fig.~\ref{fig:CWP}(a) captures the channel-specific CWPs at $c=c^{o/d}_{\mathrm{crit}}$ and $\Delta\phi=0$, where the dark-fringe working point is attached to the bare resonance ($\omega=\Omega_0, \gamma=0$). Tuning the system to the multimode exceptional point ($c=c^{\mathrm{MEP}}=1, \Delta\phi=0$)
induces
a simultaneous dark-fringe working point in both detector configurations at $\gamma=\Gamma$ [Fig.~\ref{fig:CWP}(e)]. Varying the optical phase difference $\Delta\phi$  migrates these dark-fringe working points around the central multimode exceptional point pole as illustrated in  Fig.~\ref{fig:CWP}(c). This complex-plane tracking visually confirms that at exactly $\Delta\phi=\pi$, the off-diagonal dark-fringe working points run off to infinity [Fig.~\ref{fig:topological_evolution}(c) in the main text], whereas the diagonal zeros bypass the pole and maintain finite, well-defined topological trajectories.

Fig.~\ref{fig:finite_loss} confirms the structural stability of the working points of the exceptional interferometer under non-ideal operating conditions by introducing a finite background loss ($\gamma=0.01$). This analysis replicates the primary interference landscape of Fig.~\ref{fig:main_results} in the main text. While the addition of finite dissipation induces a baseline shift in the exact spectral coordinates of the interference fringes, the global non-Hermitian topology remains  intact. The quantized winding numbers $\mathcal{W}$ continue to exhibit perfectly discrete plateaus, and abrupt topological phase transitions at the shifted values
\begin{equation}
\begin{aligned}
\frac{\kappa_2}{\kappa_1}&=\left(1-\frac{\Gamma^2-\gamma^2}{2\Gamma J}\right)^2   && \quad\mbox{(diagonal)},
\\
\frac{\kappa_2}{\kappa_1}&=\frac{1-\frac{\Gamma^2-\gamma^2}{2\Gamma J}}{1+\frac{\Gamma}{2J}}    && \quad\mbox{(off-diagonal)}
,
\end{aligned}
\label{eq:critsup}
\end{equation}
dictate the emergence and annihilation of dark-fringe pairs determined by $\mathcal{Z}=4-2\mathcal{W}$.
This confirms that our proposed non-Hermitian interferometry paradigm, with its distinct bright and dark fringe operating regimes, extends to uniform background losses. Consequently, this fundamental structural robustness ensures the experimental viability of the exceptional interferometer, firmly establishing it as a resilient architecture for next-generation precision metrology.

\bibliography{refs2.bib}

@article{young,
    author = {Young, Thomas},
    title = {I. {The} {Bakerian} Lecture. {Experiments} and calculations relative to physical optics},
    journal = {Philos. Trans. Royal Soc.},
    volume={(94)},
    number = {94},
    pages = {1-16},
    year = {1804},
    month = {12},
    issn = {0261-0523},
    doi = {10.1098/rstl.1804.0001},
    nourl = {https://doi.org/10.1098/rstl.1804.0001},
    noeprint = {https://royalsocietypublishing.org/rstl/article-pdf/doi/10.1098/rstl.1804.0001/1463503/rstl.1804.0001.pdf},
    notes={checked}
}

@article{fresnel1816memoire,
  title={M{\'e}moire sur la diffraction de la lumi{\`e}re, o{\`u} l'on examine particuli{\`e}rement le ph{\'e}nom{\`e}nes des franges color{\'e}es que pr{\'e}sentent les ombres des corps {\'e}clair{\'e}s par un point lumineux  ({Memoir} on the diffraction of light, where we especially examine the phenomenon of coloured fringes shown by the shadows of bodies illuminated by a bright point)},
  shorttitle={M{\'e}moire sur la diffraction de la lumi{\`e}re}, 
  author={Fresnel, Augustin Jean},
  journal={Ann. Chim. Phys. 2nd Ser.},
  year={1816},
  volume={1},
  pages = {239},
  url={https://gallica.bnf.fr/ark:/12148/bpt6k6570847h/f249.item},
notes={
checked
}
}

@article{arago1819action,
  title={Mémoire sur l'action que les rayons de lumière polarisée exercent les uns sur les autres ({On} the action of rays of polarized light upon each other)},
  author={Arago, D. F. J. and Fresnel, A. J.},
  journal={Ann. Chim.  Phys. 2nd Ser.},
  volume={10},
  pages={288--305},
  year={1819},
  url={https://gallica.bnf.fr/ark:/12148/bpt6k6570892b/f294.item},
notes={checked}
}

@article{Michelson1887On,
	author = {Michelson, A. A. and Morley, E. W.},
	journal = {American Journal of Science},
	doi = {10.2475/ajs.s3-34.203.333},
	number = {203},
	year = {1887},
	month = {nov 1},
	pages = {333--345},
	publisher = {Am. J. Sci.},
	title = {On the relative motion of the {Earth} and the luminiferous ether},
	volume = {s3-34},
      notes={checked}
}

@book{nolte2023,
    author = {Nolte, David D.},
    title = {Interference: The History of Optical Interferometry and the Scientists Who Tamed Light},
    publisher = {Oxford University Press},
    year = {2023},
    month = {07},
    isbn = {9780192869760},
    doi = {10.1093/oso/9780192869760.001.0001},
    url = {https://doi.org/10.1093/oso/9780192869760.001.0001},
notes={checked}
}

@book{hecht2017optics,
  title={Optics},
  author={Hecht, E.},
  isbn={9780133977226},
  lccn={2016000110},
  url={https://books.google.co.uk/books?id=ZarLoQEACAAJ},
  year={2017},
  publisher={Pearson Education},
  notes={checked}
}

@book{scully2014quantum,
  title={Quantum Optics},
  author={Scully, M.O. and Zubairy, M.S.},
  isbn={9781299773714},
  url={https://books.google.co.uk/books?id=oX3xoAEACAAJ},
  year={2014},
  publisher={Cambridge University Press},
  notes={checked}
}

@book{meystre2014elements,
  title={Elements of Quantum Optics},
  author={Meystre, P. and Sargent, M.},
  isbn={9783662038789},
  url={https://books.google.co.uk/books?id=LqsEswEACAAJ},
  year={2014},
  publisher={Springer Berlin Heidelberg},
  notes={checked}
}

@article{ligo,
doi = {10.1088/0034-4885/72/7/076901},
url = {https://doi.org/10.1088/0034-4885/72/7/076901},
year = {2009},
month = {jun},
publisher = {},
volume = {72},
number = {7},
pages = {076901},
author = {Abbott  {{et al.}}, B P },
allauthor = {Abbott, B P and Abbott, R and Adhikari, R and Ajith, P and Allen, B and Allen, G and Amin, R S and Anderson, S B and Anderson, W G and Arain, M A and Araya, M and Armandula, H and Armor, P and Aso, Y and Aston, S and Aufmuth, P and Aulbert, C and Babak, S and Baker, P and Ballmer, S and Barker, C and Barker, D and Barr, B and Barriga, P and Barsotti, L and Barton, M A and Bartos, I and Bassiri, R and Bastarrika, M and Behnke, B and Benacquista, M and Betzwieser, J and Beyersdorf, P T and Bilenko, I A and Billingsley, G and Biswas, R and Black, E and Blackburn, J K and Blackburn, L and Blair, D and Bland, B and Bodiya, T P and Bogue, L and Bork, R and Boschi, V and Bose, S and Brady, P R and Braginsky, V B and Brau, J E and Bridges, D O and Brinkmann, M and Brooks, A F and Brown, D A and Brummit, A and Brunet, G and Bullington, A and Buonanno, A and Burmeister, O and Byer, R L and Cadonati, L and Camp, J B and Cannizzo, J and Cannon, K C and Cao, J and Cardenas, L and Caride, S and Castaldi, G and Caudill, S and Cavaglià, M and Cepeda, C and Chalermsongsak, T and Chalkley, E and Charlton, P and Chatterji, S and Chelkowski, S and Chen, Y and Christensen, N and Chung, C T Y and Clark, D and Clark, J and Clayton, J H and Cokelaer, T and Colacino, C N and Conte, R and Cook, D and Corbitt, T R C and Cornish, N and Coward, D and Coyne, D C and Creighton, J D E and Creighton, T D and Cruise, A M and Culter, R M and Cumming, A and Cunningham, L and Danilishin, S L and Danzmann, K and Daudert, B and Davies, G and Daw, E J and DeBra, D and Degallaix, J and Dergachev, V and Desai, S and DeSalvo, R and Dhurandhar, S and Díaz, M and Dietz, A and Donovan, F and Dooley, K L and Doomes, E E and Drever, R W P and Dueck, J and Duke, I and Dumas, J-C and Dwyer, J G and Echols, C and Edgar, M and Effler, A and Ehrens, P and Espinoza, E and Etzel, T and Evans, M and Evans, T and Fairhurst, S and Faltas, Y and Fan, Y and Fazi, D and Fehrmenn, H and Finn, L S and Flasch, K and Foley, S and Forrest, C and Fotopoulos, N and Franzen, A and Frede, M and Frei, M and Frei, Z and Freise, A and Frey, R and Fricke, T and Fritschel, P and Frolov, V V and Fyffe, M and Galdi, V and Garofoli, J A and Gholami, I and Giaime, J A and Giampanis, S and Giardina, K D and Goda, K and Goetz, E and Goggin, L M and González, G and Gorodetsky, M L and Goßler, S and Gouaty, R and Grant, A and Gras, S and Gray, C and Gray, M and Greenhalgh, R J S and Gretarsson, A M and Grimaldi, F and Grosso, R and Grote, H and Grunewald, S and Guenther, M and Gustafson, E K and Gustafson, R and Hage, B and Hallam, J M and Hammer, D and Hammond, G D and Hanna, C and Hanson, J and Harms, J and Harry, G M and Harry, I W and Harstad, E D and Haughian, K and Hayama, K and Heefner, J and Heng, I S and Heptonstall, A and Hewitson, M and Hild, S and Hirose, E and Hoak, D and Hodge, K A and Holt, K and Hosken, D J and Hough, J and Hoyland, D and Hughey, B and Huttner, S H and Ingram, D R and Isogai, T and Ito, M and Ivanov, A and Johnson, B and Johnson, W W and Jones, D I and Jones, G and Jones, R and Ju, L and Kalmus, P and Kalogera, V and Kandhasamy, S and Kanner, J and Kasprzyk, D and Katsavounidis, E and Kawabe, K and Kawamura, S and Kawazoe, F and Kells, W and Keppel, D G and Khalaidovski, A and Khalili, F Y and Khan, R and Khazanov, E and King, P and Kissel, J S and Klimenko, S and Kokeyama, K and Kondrashov, V and Kopparapu, R and Koranda, S and Kozak, D and Krishnan, B and Kumar, R and Kwee, P and Lam, P K and Landry, M and Lantz, B and Lazzarini, A and Lei, H and Lei, M and Leindecker, N and Leonor, I and Li, C and Lin, H and Lindquist, P E and Littenberg, T B and Lockerbie, N A and Lodhia, D and Longo, M and Lormand, M and Lu, P and Lubinski, M and Lucianetti, A and Lück, H and Machenschalk, B and MacInnis, M and Mageswaran, M and Mailand, K and Mandel, I and Mandic, V and Márka, S and Márka, Z and Markosyan, A and Markowitz, J and Maros, E and Martin, I W and Martin, R M and Marx, J N and Mason, K and Matichard, F and Matone, L and Matzner, R A and Mavalvala, N and McCarthy, R and McClelland, D E and McGuire, S C and McHugh, M and McIntyre, G and McKechan, D J A and McKenzie, K and Mehmet, M and Melatos, A and Melissinos, A C and Menéndez, D F and Mendell, G and Mercer, R A and Meshkov, S and Messenger, C and Meyer, M S and Miller, J and Minelli, J and Mino, Y and Mitrofanov, V P and Mitselmakher, G and Mittleman, R and Miyakawa, O and Moe, B and Mohanty, S D and Mohapatra, S R P and Moreno, G and Morioka, T and Mors, K and Mossavi, K and MowLowry, C and Mueller, G and Müller-Ebhardt, H and Muhammad, D and Mukherjee, S and Mukhopadhyay, H and Mullavey, A and Munch, J and Murray, P G and Myers, E and Myers, J and Nash, T and Nelson, J and Newton, G and Nishizawa, A and Numata, K and O'Dell, J and O'Reilly, B and O'Shaughnessy, R and Ochsner, E and Ogin, G H and Ottaway, D J and Ottens, R S and Overmier, H and Owen, B J and Pan, Y and Pankow, C and Papa, M A and Parameshwaraiah, V and Patel, P and Pedraza, M and Penn, S and Perraca, A and Pierro, V and Pinto, I M and Pitkin, M and Pletsch, H J and Plissi, M V and Postiglione, F and Principe, M and Prix, R and Prokhorov, L and Punken, O and Quetschke, V and Raab, F J and Rabeling, D S and Radkins, H and Raffai, P and Raics, Z and Rainer, N and Rakhmanov, M and Raymond, V and Reed, C M and Reed, T and Rehbein, H and Reid, S and Reitze, D H and Riesen, R and Riles, K and Rivera, B and Roberts, P and Robertson, N A and Robinson, C and Robinson, E L and Roddy, S and Röver, C and Rollins, J and Romano, J D and Romie, J H and Rowan, S and Rüdiger, A and Russell, P and Ryan, K and Sakata, S and de la Jordana, L Sancho and Sandberg, V and Sannibale, V and Santamaría, L and Saraf, S and Sarin, P and Sathyaprakash, B S and Sato, S and Satterthwaite, M and Saulson, P R and Savage, R and Savov, P and Scanlan, M and Schilling, R and Schnabel, R and Schofield, R and Schulz, B and Schutz, B F and Schwinberg, P and Scott, J and Scott, S M and Searle, A C and Sears, B and Seifert, F and Sellers, D and Sengupta, A S and Sergeev, A and Shapiro, B and Shawhan, P and Shoemaker, D H and Sibley, A and Siemens, X and Sigg, D and Sinha, S and Sintes, A M and Slagmolen, B J J and Slutsky, J and Smith, J R and Smith, M R and Smith, N D and Somiya, K and Sorazu, B and Stein, A and Stein, L C and Steplewski, S and Stochino, A and Stone, R and Strain, K A and Strigin, S and Stroeer, A and Stuver, A L and Summerscales, T Z and Sun, K-X and Sung, M and Sutton, P J and Szokoly, G P and Talukder, D and Tang, L and Tanner, D B and Tarabrin, S P and Taylor, J R and Taylor, R and Thacker, J and Thorne, K A and Thüring, A and Tokmakov, K V and Torres, C and Torrie, C and Traylor, G and Trias, M and Ugolini, D and Ulmen, J and Urbanek, K and Vahlbruch, H and Vallisneri, M and Broeck, C Van Den and van der Sluys, M V and van Veggel, A A and Vass, S and Vaulin, R and Vecchio, A and Veitch, J and Veitch, P and Veltkamp, C and Villar, A and Vorvick, C and Vyachanin, S P and Waldman, S J and Wallace, L and Ward, R L and Weidner, A and Weinert, M and Weinstein, A J and Weiss, R and Wen, L and Wen, S and Wette, K and Whelan, J T and Whitcomb, S E and Whiting, B F and Wilkinson, C and Willems, P A and Williams, H R and Williams, L and Willke, B and Wilmut, I and Winkelmann, L and Winkler, W and Wipf, C C and Wiseman, A G and Woan, G and Wooley, R and Worden, J and Wu, W and Yakushin, I and Yamamoto, H and Yan, Z and Yoshida, S and Zanolin, M and Zhang, J and Zhang, L and Zhao, C and Zotov, N and Zucker, M E and Mühlen, H zur and Zweizig, J},
title = {{LIGO}: the Laser Interferometer Gravitational-Wave Observatory},
journal = {Rep. Prog. Phys.},
notes={checked}
}

@article{microjan2,
  title = {Dielectric microcavities: Model systems for wave chaos and non-Hermitian physics},
  author = {Cao, Hui and Wiersig, Jan},
  journal = {Rev. Mod. Phys.},
  volume = {87},
  issue = {1},
  pages = {61--111},
  numpages = {51},
  year = {2015},
  month = {Jan},
  publisher = {American Physical Society},
  doi = {10.1103/RevModPhys.87.61},
  url = {https://link.aps.org/doi/10.1103/RevModPhys.87.61},
  notes={checked}
}

@Article{ElGanainy2018,
author={El-Ganainy, Ramy
and Makris, Konstantinos G.
and Khajavikhan, Mercedeh
and Musslimani, Ziad H.
and Rotter, Stefan
and Christodoulides, Demetrios N.},
title={Non-{Hermitian} physics and {PT} symmetry},
journal={Nat. Phys.},
year={2018},
month={Jan},
day={01},
volume={14},
number={1},
pages={11-19},
issn={1745-2481},
doi={10.1038/nphys4323},
url={https://doi.org/10.1038/nphys4323},
  notes={checked}
}

@BOOK{Kato66,
    AUTHOR    = "Kato, T.",
    TITLE     = "Perturbation Theory for Linear Operators",
    PUBLISHER = "Springer",
    ADDRESS   = "New York",
    DOI       = {10.1007/978-3-642-66282-9},
    URL       = {https://www.springer.com/de/book/9783540586616},
    YEAR      = 1966,
  notes={fromjan}
}

@article{Heiss2012,
doi = {10.1088/1751-8113/45/44/444016},
url = {https://doi.org/10.1088/1751-8113/45/44/444016},
year = {2012},
month = {oct},
publisher = {IOP Publishing},
volume = {45},
number = {44},
pages = {444016},
author = {Heiss, W D},
title = {The physics of exceptional points},
journal = {J. Phys. A},
  notes={checked}
}

@article{Miri2019,
  title={Exceptional points in optics and photonics},
  author={Miri, Mohammad-Ali and  Al\`{u}, Andrea},
  journal={Science},
  volume={363},
  number={6422},
  pages={eaar7709},
  year={2019},
  DOI     = {10.1126/science.aar7709},
 URL     = {http://dx.doi.org/10.1126/science.aar7709},
  publisher={American Association for the Advancement of Science},
  notes={checked}
}

@article{Dembowski2001,
  title = {Experimental Observation of the Topological Structure of Exceptional Points},
  author = {Dembowski, C. and Gr\"af, H.-D. and Harney, H. L. and Heine, A. and Heiss, W. D. and Rehfeld, H. and Richter, A.},
  journal = {Phys. Rev. Lett.},
  volume = {86},
  issue = {5},
  pages = {787--790},
  numpages = {0},
  year = {2001},
  month = {Jan},
  publisher = {American Physical Society},
  doi = {10.1103/PhysRevLett.86.787},
  url = {https://link.aps.org/doi/10.1103/PhysRevLett.86.787},
notes={checked}
}

@Article{Berry2004,
author={Berry, M. V.},
title={Physics of Nonhermitian Degeneracies},
journal={Czech. J. Phys.},
year={2004},
month={Oct},
day={01},
volume={54},
number={10},
pages={1039-1047},
issn={1572-9486},
doi={10.1023/B:CJOP.0000044002.05657.04},
url={https://doi.org/10.1023/B:CJOP.0000044002.05657.04},
notes={checked}
}

@Article{Doppler2016,
author={Doppler, J{\"o}rg
and Mailybaev, Alexei A.
and B{\"o}hm, Julian
and Kuhl, Ulrich
and Girschik, Adrian
and Libisch, Florian
and Milburn, Thomas J.
and Rabl, Peter
and Moiseyev, Nimrod
and Rotter, Stefan},
title={Dynamically encircling an exceptional point for asymmetric mode switching},
journal={Nature},
year={2016},
month={Sep},
day={01},
volume={537},
number={7618},
pages={76-79},
issn={1476-4687},
doi={10.1038/nature18605},
url={https://doi.org/10.1038/nature18605},
notes={checked}
}

@Article{Ghosh2016,
author={Ghosh, S. N.
and Chong, Y. D.},
title={Exceptional points and asymmetric mode conversion in quasi-guided dual-mode optical waveguides},
journal={Sci. Rep.},
year={2016},
month={Apr},
day={22},
volume={6},
number={1},
pages={19837},
issn={2045-2322},
doi={10.1038/srep19837},
url={https://doi.org/10.1038/srep19837},
notes={checked}
}

@article{microjan1,
  title = {Enhancing the Sensitivity of Frequency and Energy Splitting Detection by Using Exceptional Points: Application to Microcavity Sensors for Single-Particle Detection},
  author = {Wiersig, Jan},
  journal = {Phys. Rev. Lett.},
  volume = {112},
  issue = {20},
  pages = {203901},
  numpages = {5},
  year = {2014},
  month = {May},
  publisher = {American Physical Society},
  doi = {10.1103/PhysRevLett.112.203901},
  url = {https://link.aps.org/doi/10.1103/PhysRevLett.112.203901},
notes={checked}
}

@Article{Chen2017,
author={Chen, Weijian
and  {\"O}zdemir, {\c{S}}ahin Kaya
and Zhao, Guangming
and Wiersig, Jan
and Yang, Lan},
title={Exceptional points enhance sensing in an optical microcavity},
journal={Nature},
year={2017},
month={Aug},
day={01},
volume={548},
number={7666},
pages={192-196},
issn={1476-4687},
doi={10.1038/nature23281},
url={https://doi.org/10.1038/nature23281},
notes={checked}
}

@Article{Hodaei2017,
author={Hodaei, Hossein
and Hassan, Absar U.
and Wittek, Steffen
and Garcia-Gracia, Hipolito
and El-Ganainy, Ramy
and Christodoulides, Demetrios N.
and Khajavikhan, Mercedeh},
title={Enhanced sensitivity at higher-order exceptional points},
journal={Nature},
year={2017},
month={Aug},
day={01},
volume={548},
number={7666},
pages={187-191},
issn={1476-4687},
doi={10.1038/nature23280},
url={https://doi.org/10.1038/nature23280},
notes={checked}
}

@article{Wiersig:20,
author = {Jan Wiersig},
journal = {Photon. Res.},
number = {9},
pages = {1457--1467},
publisher = {Optica Publishing Group},
title = {Review of exceptional point-based sensors},
volume = {8},
month = {Sep},
year = {2020},
url = {https://opg.optica.org/prj/abstract.cfm?URI=prj-8-9-1457},
doi = {10.1364/PRJ.396115},
notes={checked}
}

@Article{Peng2014,
author={Peng, Bo
and {\"O}zdemir, {\c{S}}ahin Kaya
and Lei, Fuchuan
and Monifi, Faraz
and Gianfreda, Mariagiovanna
and Long, Gui Lu
and Fan, Shanhui
and Nori, Franco
and Bender, Carl M.
and Yang, Lan},
title={Parity--time-symmetric whispering-gallery microcavities},
journal={Nat. Phys.},
year={2014},
month={May},
day={01},
volume={10},
number={5},
pages={394-398},
issn={1745-2481},
doi={10.1038/nphys2927},
url={https://doi.org/10.1038/nphys2927},
notes={checked}
}

@Article{Xu2016,
author={Xu, H.
and Mason, D.
and Jiang, Luyao
and Harris, J. G. E.},
title={Topological energy transfer in an optomechanical system with exceptional points},
journal={Nature},
year={2016},
month={Sep},
day={01},
volume={537},
number={7618},
pages={80-83},
issn={1476-4687},
doi={10.1038/nature18604},
url={https://doi.org/10.1038/nature18604},
notes={checked}
}

@Article{Zhang2017,
author={Zhang, Dengke
and Luo, Xiao-Qing
and Wang, Yi-Pu
and Li, Tie-Fu
and You, J. Q.},
title={Observation of the exceptional point in cavity magnon-polaritons},
journal={Nat. Commun.},
year={2017},
month={Nov},
day={08},
volume={8},
number={1},
pages={1368},
issn={2041-1723},
doi={10.1038/s41467-017-01634-w},
url={https://doi.org/10.1038/s41467-017-01634-w},
notes={checked}
}

@article{Zhong2019,
  title = {Sensing with Exceptional Surfaces in Order to Combine Sensitivity with Robustness},
  author = {Zhong, Q. and Ren, J. and Khajavikhan, M. and Christodoulides, D. N. and \"Ozdemir, {\c{S}}. K. and El-Ganainy, R.},
  journal = {Phys. Rev. Lett.},
  volume = {122},
  issue = {15},
  pages = {153902},
  numpages = {6},
  year = {2019},
  month = {Apr},
  publisher = {American Physical Society},
  doi = {10.1103/PhysRevLett.122.153902},
  url = {https://link.aps.org/doi/10.1103/PhysRevLett.122.153902},
notes={checked}
}

@Article{Li2023,
author={Li, Aodong
and Wei, Heng
and Cotrufo, Michele
and Chen, Weijin
and Mann, Sander
and Ni, Xiang
and Xu, Bingcong
and Chen, Jianfeng
and Wang, Jian
and Fan, Shanhui
and Qiu, Cheng-Wei
and Al{\`u}, Andrea
and Chen, Lin},
title={Exceptional points and non-{Hermitian} photonics at the nanoscale},
journal={Nat. Nanotechnol.},
year={2023},
month={Jul},
day={01},
volume={18},
number={7},
pages={706-720},
issn={1748-3395},
doi={10.1038/s41565-023-01408-0},
url={https://doi.org/10.1038/s41565-023-01408-0}
}

@article{Zhong2020,
  title = {Hierarchical Construction of Higher-Order Exceptional Points},
author = {Zhong, Q. and Kou, J. and \"Ozdemir, {\c{S}}. K. and El-Ganainy, R.},
  journal = {Phys. Rev. Lett.},
  volume = {125},
  issue = {20},
  pages = {203602},
  numpages = {5},
  year = {2020},
  month = {Nov},
  publisher = {American Physical Society},
  doi = {10.1103/PhysRevLett.125.203602},
  url = {https://link.aps.org/doi/10.1103/PhysRevLett.125.203602},
notes={checked}
}

@Article{Hashemi2022,
author={Hashemi, A.
and Busch, K.
and Christodoulides, D. N.
and Ozdemir, S. K.
and El-Ganainy, R.},
title={Linear response theory of open systems with exceptional points},
journal={Nat. Commun.},
year={2022},
month={Jun},
day={07},
volume={13},
number={1},
pages={3281},
issn={2041-1723},
doi={10.1038/s41467-022-30715-8},
url={https://doi.org/10.1038/s41467-022-30715-8},
notes={checked}
}

@article{Kullig:23,
author = {Julius Kullig and Daniel Grom and Sebastian Klembt and Jan Wiersig},
journal = {Photon. Res.},
number = {10},
pages = {A54--A64},
publisher = {Optica Publishing Group},
title = {Higher-order exceptional points in waveguide-coupled microcavities: perturbation induced frequency splitting and mode patterns},
volume = {11},
month = {Oct},
year = {2023},
url = {https://opg.optica.org/prj/abstract.cfm?URI=prj-11-10-A54},
doi = {10.1364/PRJ.496414},
notes={checked}
}

@book{Loudon,
    author = {Loudon, Rodney},
    title = {The Quantum Theory of Light},
    publisher = {Oxford University Press},
    year = {2000},
    month = {09},
    isbn = {9780198501770},
    doi = {10.1093/oso/9780198501770.001.0001},
    url = {https://doi.org/10.1093/oso/9780198501770.001.0001},
notes={checked}
}

@Article{Ruter2010,
author={R{\"u}ter, Christian E.
and Makris, Konstantinos G.
and El-Ganainy, Ramy
and Christodoulides, Demetrios N.
and Segev, Mordechai
and Kip, Detlef},
title={Observation of parity--time symmetry in optics},
journal={Nat. Phys.},
year={2010},
month={Mar},
day={01},
volume={6},
number={3},
pages={192-195},
issn={1745-2481},
doi={10.1038/nphys1515},
url={https://doi.org/10.1038/nphys1515},
notes={checked}
}

@article{QuirozJuarez:19,
author = {Mario A. Quiroz-Ju\'{a}rez and Armando Perez-Leija and Konrad Tschernig and Blas M. Rodr\'{i}guez-Lara and Omar S. {Maga\~{n}a-Loaiza} and Kurt Busch and Yogesh N. Joglekar and Roberto de J. Le\'{o}n-Montiel},
journal = {Photon. Res.},
number = {8},
pages = {862--867},
publisher = {Optica Publishing Group},
title = {Exceptional points of any order in a single, lossy waveguide beam splitter by photon-number-resolved detection},
volume = {7},
month = {Aug},
year = {2019},
url = {https://opg.optica.org/prj/abstract.cfm?URI=prj-7-8-862},
doi = {10.1364/PRJ.7.000862},
notes={checked}
}

@Article{On2024,
author={On, Mehmet Berkay
and Ashtiani, Farshid
and Sanchez-Jacome, David
and Perez-Lopez, Daniel
and Yoo, S. J. Ben
and Blanco-Redondo, Andrea},
title={Programmable integrated photonics for topological {Hamiltonians}},
journal={Nat. Commun.},
year={2024},
month={Jan},
day={20},
volume={15},
number={1},
pages={629},
issn={2041-1723},
doi={10.1038/s41467-024-44939-3},
url={https://doi.org/10.1038/s41467-024-44939-3},
notes={checked}
}

@book{Van2016Microring,
  author    = {Van, V.},
  title     = {Optical Microring Resonators: Theory, Techniques, and Applications},
  edition   = {1st},
  year      = {2016},
  publisher = {CRC Press},
  doi       = {10.1201/9781315303512}
}

@Article{Soleymani2022,
author={Soleymani, S.
and Zhong, Q.
and Mokim, M.
and Rotter, S.
and El-Ganainy, R.
and {\"O}zdemir, {\c{S}}. K.},
title={Chiral and degenerate perfect absorption on exceptional surfaces},
journal={Nat. Commun.},
year={2022},
month={Feb},
day={01},
volume={13},
number={1},
pages={599},
issn={2041-1723},
doi={10.1038/s41467-022-27990-w},
url={https://doi.org/10.1038/s41467-022-27990-w},
notes={checked} 
}

@article{Schindler2012,
doi = {10.1088/1751-8113/45/44/444029},
url = {https://doi.org/10.1088/1751-8113/45/44/444029},
year = {2012},
month = {oct},
publisher = {IOP Publishing},
volume = {45},
number = {44},
pages = {444029},
author = {Schindler, J and Lin, Z and Lee, J M and Ramezani, H and Ellis, F M and Kottos, T},
title = {$\mathcal{PT}$-symmetric electronics},
journal = {J. Phys. A},
notes={checked} 
}

@article{zhu2014pt,
  title = {$\mathcal{P}\mathcal{T}$-Symmetric Acoustics},
  author = {Zhu, Xuefeng and Ramezani, Hamidreza and Shi, Chengzhi and Zhu, Jie and Zhang, Xiang},
  journal = {Phys. Rev. X},
  volume = {4},
  issue = {3},
  pages = {031042},
  numpages = {7},
  year = {2014},
  month = {Sep},
  publisher = {American Physical Society},
  doi = {10.1103/PhysRevX.4.031042},
  url = {https://link.aps.org/doi/10.1103/PhysRevX.4.031042},
notes={checked} 
}

@article{Yoo2011,
  title = {Quantum noise and mode nonorthogonality in non-{Hermitian} $\mathcal{PT}$-symmetric optical resonators},
  author = {Yoo, Gwangsu and Sim, H.-S. and Schomerus, Henning},
  journal = {Phys. Rev. A},
  volume = {84},
  issue = {6},
  pages = {063833},
  numpages = {14},
  year = {2011},
  month = {Dec},
  publisher = {American Physical Society},
  doi = {10.1103/PhysRevA.84.063833},
  url = {https://link.aps.org/doi/10.1103/PhysRevA.84.063833},
notes={checked} 
}

@article{Bergholtz2021,
  title = {Exceptional topology of non-Hermitian systems},
  author = {Bergholtz, Emil J. and Budich, Jan Carl and Kunst, Flore K.},
  journal = {Rev. Mod. Phys.},
  volume = {93},
  issue = {1},
  pages = {015005},
  numpages = {31},
  year = {2021},
  month = {Feb},
  publisher = {American Physical Society},
  doi = {10.1103/RevModPhys.93.015005},
  url = {https://link.aps.org/doi/10.1103/RevModPhys.93.015005},
notes={checked} 
}

@article{Kawabata2019,
  title = {Symmetry and Topology in Non-Hermitian Physics},
  author = {Kawabata, Kohei and Shiozaki, Ken and Ueda, Masahito and Sato, Masatoshi},
  journal = {Phys. Rev. X},
  volume = {9},
  issue = {4},
  pages = {041015},
  numpages = {52},
  year = {2019},
  month = {Oct},
  publisher = {American Physical Society},
  doi = {10.1103/PhysRevX.9.041015},
  url = {https://link.aps.org/doi/10.1103/PhysRevX.9.041015},
notes={checked} 
}

@article{Langbein2018,
  title = {No exceptional precision of exceptional-point sensors},
  author = {Langbein, W.},
  journal = {Phys. Rev. A},
  volume = {98},
  issue = {2},
  pages = {023805},
  numpages = {8},
  year = {2018},
  month = {Aug},
  publisher = {American Physical Society},
  doi = {10.1103/PhysRevA.98.023805},
  url = {https://link.aps.org/doi/10.1103/PhysRevA.98.023805},
notes={checked} 
}

@article{Caves1981,
  title = {Quantum-mechanical noise in an interferometer},
  author = {Caves, Carlton M.},
  journal = {Phys. Rev. D},
  volume = {23},
  issue = {8},
  pages = {1693--1708},
  numpages = {0},
  year = {1981},
  month = {Apr},
  publisher = {American Physical Society},
  doi = {10.1103/PhysRevD.23.1693},
  url = {https://link.aps.org/doi/10.1103/PhysRevD.23.1693},
notes={checked} 
}

@Article{Naghiloo2019,
author={Naghiloo, M.
and Abbasi, M.
and Joglekar, Yogesh N.
and Murch, K. W.},
title={Quantum state tomography across the exceptional point in a single dissipative qubit},
journal={Nat. Phys.},
year={2019},
month={Dec},
day={01},
volume={15},
number={12},
pages={1232-1236},
issn={1745-2481},
doi={10.1038/s41567-019-0652-z},
url={https://doi.org/10.1038/s41567-019-0652-z},
notes={checked} 
}

@Article{Heiss2015,
author={Heiss, W. D.},
title={Green's Functions at Exceptional Points},
journal={Int. J. Theor. Phys.},
year={2015},
month={Nov},
day={01},
volume={54},
number={11},
pages={3954-3959},
issn={1572-9575},
doi={10.1007/s10773-014-2428-7},
url={https://doi.org/10.1007/s10773-014-2428-7},
notes={checked} 
}

@article{Wiersig2023,
  title = {Moving along an exceptional surface towards a higher-order exceptional point},
  author = {Wiersig, Jan},
  journal = {Phys. Rev. A},
  volume = {108},
  issue = {3},
  pages = {033501},
  numpages = {10},
  year = {2023},
  month = {Sep},
  publisher = {American Physical Society},
  doi = {10.1103/PhysRevA.108.033501},
  url = {https://link.aps.org/doi/10.1103/PhysRevA.108.033501},
notes={checked} 
}

@article{bid2024uniform,
  title = {Uniform response theory of non-{Hermitian} systems: Non-{Hermitian} physics beyond the exceptional point},
  author = {Bid, Subhajyoti and Schomerus, Henning},
  journal = {Phys. Rev. Res.},
  volume = {7},
  issue = {2},
  pages = {023062},
  numpages = {24},
  year = {2025},
  month = {Apr},
  publisher = {American Physical Society},
  doi = {10.1103/PhysRevResearch.7.023062},
  url = {https://link.aps.org/doi/10.1103/PhysRevResearch.7.023062},
notes={checked} 
}

@article{Fan:03,
author = {Shanhui Fan and Wonjoo Suh and J. D. Joannopoulos},
journal = {J. Opt. Soc. Am. A},
number = {3},
pages = {569--572},
publisher = {Optica Publishing Group},
title = {Temporal coupled-mode theory for the {Fano} resonance in optical resonators},
volume = {20},
month = {Mar},
year = {2003},
url = {https://opg.optica.org/josaa/abstract.cfm?URI=josaa-20-3-569},
doi = {10.1364/JOSAA.20.000569},
notes={checked} 
}

@article{researchdata,
title = "Research data for ``{Non}-{Hermitian} multimode interferometry'' {[Dataset and software]}",
journal={Zenodo},
 author = {Bid, Subhajyoti and Schomerus, Henning},
 year={2026},
 url={https://doi.org/10.5281/zenodo.21872470},
  doi={https://doi.org/10.5281/zenodo.21872470}
 }

\end{document}